# Coupled Mesoscale-Large-Eddy Modeling of Realistic Stable Boundary Layer Turbulence

Yao Wang,[1] Sukanta Basu,[1, a)] and Lance Manuel[2]

[1)] *Department of Marine, Earth, and Atmospheric Sciences, North Carolina State University, Raleigh, NC 27695*

[2)] *Department of Civil, Architectural and Environmental Engineering, University of Texas, Austin, USA*

Site-specific flow and turbulence information are needed for various practical applications, ranging from aerodynamic/aeroelastic modeling for wind turbine design to optical diffraction calculations. Even though highly desirable, collecting on-site meteorological measurements can be an expensive, time-consuming, and sometimes a challenging task. In this work, we propose a coupled mesoscale-large-eddy modeling framework to synthetically generate site-specific flow and turbulence data. The workhorses behind our framework are a state-of-the-art, open-source atmospheric model called the Weather Research and Forecasting (WRF) model and a tuning-free large-eddy simulation (LES) model.

Using this coupled framework, we simulate a nighttime stable boundary layer (SBL) case from the well-known CASES-99 field campaign. One of the unique aspects of this work is the usage of a diverse range of observations for characterization and validation. The coupled models reproduce certain characteristics of observed low-level jets. They also capture various scaling regimes of energy spectra, including the so-called spectral gap. However, the coupled models are unable to capture the intermittent nature of the observed surface fluxes. Lastly, we document and discuss: (i) the tremendous spatio-temporal variabilities of observed and modeled SBL flow fields, and (ii) the significant disagreements among different observational platforms. Based on these results, we strongly recommend that future SBL modeling studies consider rigorous validation exercises based on multi-sensor/multi-platform datasets.

In summary, we believe that the numerical generation of realistic SBL is not an impossible task. Without any doubt, there remain several computational and fundamental challenges. The present work should be viewed as a first step to confront some of these challenges.

[a] Electronic mail: sukanta˙basu@ncsu.edu

## I. INTRODUCTION

Over the past two decades or so, several large-eddy simulation (LES) studies (e.g.,[3,8,14,21,39,45,54,75,101]) were conducted to enhance our understanding of atmospheric stable boundary layers (SBLs). Initially, some of these studies encountered various modeling challenges including, but not limited to, unusual subgrid-scale (SGS) modeling behavior, excessive resolution sensitivity, difficulty in initialization, unphysical runaway surface cooling, unexpected model crashing, etc. With the advent of robust SGS models (e.g., various types of dynamic SGS models) along with the easy availability of high-performance computing resources, the quality of SBL simulations has improved substantially. Researchers are gaining more confidence in LES-generated SBL data and have already started using them to complement sparsely available observational data. For example, recently, an extensive LES database (called the NCSU-LES database) of idealized SBLs has been utilized to address an age-old problem of stability dependence of critical bulk Richardson number[73]. Such LES studies are not only making immense contributions to the fundamental science of SBLs, they are also impacting applied studies. For example, the NCSU-LES database has also helped researchers to identify rotor-scale flow variables that influence extreme and fatigue loads on wind turbines[68].

One of the drawbacks of the aforementioned idealized LES studies is that they have fallen short with regard to capturing 'real-world' SBL turbulence. This is due to the fact that these simulations did not include: (i) the effects of natural topography and land-surface heterogeneities, (ii) baroclinicity, large-scale advection, and subsidence effects, and (iii) interactions among several physical processes—e.g., turbulence, radiative transfer, and cloud microphysics. To partially fill this void, a handful of realistic LES studies were conducted recently (e.g.,[12,13,15,46]). These studies used site-specific observations to initialize and to force their simulations. Encouraging results have been reported. For example, Basu et al.[12] showed that a tuning-free LES model can reliably capture the development, magnitude, and location of an observed nocturnal low-level jet (LLJ).

Due to a variety of logistical, financial, and instrumental issues, high-quality SBL data are difficult to collect. So, from a practical standpoint, the initialization and forcing of an LES model based on local observations has not been common. An alternative is to run a mesoscale meteorological model (MMM) for a desired geographical region and time frame and, then, to dynamically downscale the simulated fields using LES. In this study, we propose a new coupled modeling framework to achieve this goal and demonstrate its strengths and weaknesses.

The structure of this paper is as follows. In Section 2, we provide some background on existing coupled modeling frameworks. Our proposed framework is discussed in Section 3. A case study and associated observational datasets are described in Sections 4 and 5, respectively. Sections 6 and 7 provide detailed information on our modeling activities. A wide range of observed and simulated statistics are discussed in Section 8. In that section, we also document the results from several sensitivity experiments. Finally, in Section 9, we summarize our findings and provide some concluding remarks.

## II. BACKGROUND

Mesoscale meteorological models simulate or forecast atmospheric phenomena including (but not limited to) convective clouds, thunderstorms, squall lines, frontal circulations, low-level jets, terrain-induced mesoscale circulations, land-/sea-breezes, urban heat island circulations, mountain-valley winds, lee waves, and gravity waves[17,47]. The spatial and temporal scales associated with these phenomena are on the order of $\approx 2 - 2000$ km and $1 - 48$ h, respectively[17,47,67]. Over the past three decades, due to the increased availability of computational resources, the overall performance of MMMs has been steadily increasing—grid resolutions are getting finer; computational domain sizes are now larger; model physics parameterizations are becoming more complex; the number of ensemble members is growing; etc. Most importantly, due to the enhanced capabilities of the MMMs, their application arenas are also broadening. For example, even until only a decade ago, MMMs were rarely used for commercial wind energy applications. Now, they represent a significant part of the most common tools used by the wind industry (e.g.,[20]). Another recent area of increased application is in the field of laser propagation (e.g.,[52]).

Despite their versatility, MMMs cannot be used to generate high-resolution, 4-D atmospheric boundary layer (ABL) turbulence fields. This restriction is not due to any technological barriers; if anything, in the current era of petascale computing, one can utilize thousands of processors and generate flow fields with a resolution of O(10 m). However, in employing such brute-force computing, one violates two fundamental principles underlying traditional mesoscale modeling. First, MMMs solve the Reynolds-Averaged Navier Stokes (RANS) equations and, thus, the horizontal grid resolution of the MMMs must be larger than the scale of the energy- and flux-containing turbulence[97]. Since daytime ABL eddies can be as large as several hundred meters, MMMs should not be run with sub-km resolution. The second violation is related to the inherent assumption of most contemporary MMMs—that turbulence mixing is dominated by vertical mixing. To be consistent with this assumption, the aspect ratio of horizontal to vertical grid spacing near the surface should be kept at a large value (see the discussion in[100]). An aspect ratio of O(50:1) is common in practice.

Given the difficulties of running MMMs with sub-km resolution, several coupled modeling approaches have been proposed in recent studies[49,51,59,74,76,90,98,99]. In all these approaches, an MMM is either coupled with an engineering RANS model or with an LES model. These approaches can be broadly classified into two: (i) one-way coupling; and (ii) two-way coupling. Within the former class, several variants exist in the literature as is described below. Please refer to[76] and[98] for other types of classification.

## A. One-Way Coupling

In this approach, the information transfer is only one-way—from the MMM to the microscale model (MiM). This type of coupled modeling approach is relatively easy to implement. It can be used to couple two separate models or to couple the same model with different parameterizations for the MMMs and the MiMs. However, this approach faces two fundamental issues. The first issue occurs when an engineering RANS model is used as an MiM. From the literature, it is not clear whether grid resolutions utilized by the engineering RANS models are always coarser than the energy-containing eddies. If they are not, one again violates the RANS issue mentioned above. Since most of the engineering RANS models use 3-D diffusion, the aspect ratio issue mentioned above is not a problem. A different issue arises when an LES model is used as an MiM. We know that an MMM generates an 'ensemble' flow field, whereas an LES creates an 'instantaneous' field (which can be viewed as one of the member of the ensemble). So, can we really prescribe MMM-simulated values to run an LES? These issues need to be addressed in future research.

#### *1. Variant 1*

In this approach, a snapshot of the MMM run is used to provide the initial conditions for the MiM. Then, the MiM (typically a RANS model) is run to reach a steady-state condition. The MiM adjusts to the underlying fine-resolution topography or urban canopy during the simulation. Schlünzen et al.[76] referred to this approach as a 'time-slice approach'. Instead of a single snapshot, multiple snapshots (available, say, every 3 h) from an MMM run can also be used with this type of coupled approach.

#### *2. Variant 2*

In this coupled approach, the MMM provides both initial and continuous boundary conditions to the MiM—an example is the simulation reported by Baik et al.[5]. They first performed a mesoscale simulation using the MM5 model and stored the simulated results every 10 min. Then, they linearly interpolated (in time and in vertical direction) the MM5 results to provide inflow and top boundary conditions to a RANS model. A qualitatively similar approach was used by Talbot et al.[90]. They coupled the Weather Research and Forecasting (WRF) mesoscale model with the WRF-LES model.

#### *3. Variant 3*

This coupled approach is quite similar to Variant 2. Here, one uses a Newtonian relaxation method (widely known as nudging in the meteorology literature) in addition to (or sometimes in lieu of) lateral boundary conditions (e.g.,[98,99]). This approach is highly sensitive to the choice of the nudging coefficient ($G$). There is no universally accepted guidelines for the specification of $G$. For example, Yamada and Koike[98] recommended $G$ to be equal to 1% of the reciprocal of the integration time step. Thus, $G$ is equal to 0.0001 $s^{-1}$ and 0.01 $s^{-1}$, respectively, for typical MMM and MiM runs. In contrast, Zajaczkowski et al.[99] recommended that $G$ be taken to be exactly equal to the reciprocal of the integration time step (i.e., a factor of 100 different from the prescription by[98]). The need to have an optimum value of $G$ is strongly emphasized by Schlünzen et al.[76].

The Variant 3 approach suffers from a more fundamental problem, which arises due to the fact that the engineering MiMs usually do not contain important physics options (e.g., buoyancy, radiation, microphysics). To illustrate, we briefly discuss a hypothetical scenario, which is quite relevant to the present paper. Let us assume that a coupled MMM-MiM approach with nudging option is being used to simulate a nocturnal stable boundary layer. Furthermore, assume that the MiM does not contain a radiation parameterization. The MMM generates a stable boundary layer which is deep due to the presence of both turbulent and radiative flux divergences. Now, the MiM tries to capture this deep boundary layer by (spuriously) generating excessive turbulent fluxes (since it is compensating for its lack of radiation parameterization). In other words, the coupled approach would likely generate physically unrealistic flow fields.

### B. Two-Way Coupling

In this approach, the information transfer is two-way—from the MMM to the MiM and vice versa. The information transfer from the MMM to the MiM is the same as in Variant 2 of the one-way coupling approach. In addition, the aggregated MiM results are used to update the MMM results. Liu et al.[51] demonstrated the strength of this approach in short-term wind forecasting over complex terrain. It is possible that this approach suffers from 'double counting' of diffusion[99]. Based on our own research, we found that this approach occasionally suffers from numerical instabilities. We also witnessed spurious flow patterns at the nest boundaries. Similar problems were discussed in detail by Moeng[60].

## III. PROPOSED MODELING FRAMEWORK

In order to avoid the unphysical influence of the MMM results on the MiM solutions, we propose a different coupled modeling framework (see Fig. 1). In this approach, we extract initial conditions, time-dependent lower-boundary conditions (e.g., near-surface air temperature), and time-height-dependent large-scale forcings (e.g., geostrophic wind, mesoscale advection of temperature) from the MMM output and utilize them for the LES runs.

We wish to point out that our proposed framework is conceptually similar to the one recently reported by Rizza et al.[74]. However, there are some technical differences; most importantly, Rizza et al.[74] neglected the effects of mesoscale advection forcings in their study. In the present work, we will demonstrate the importance of these forcings. Furthermore,

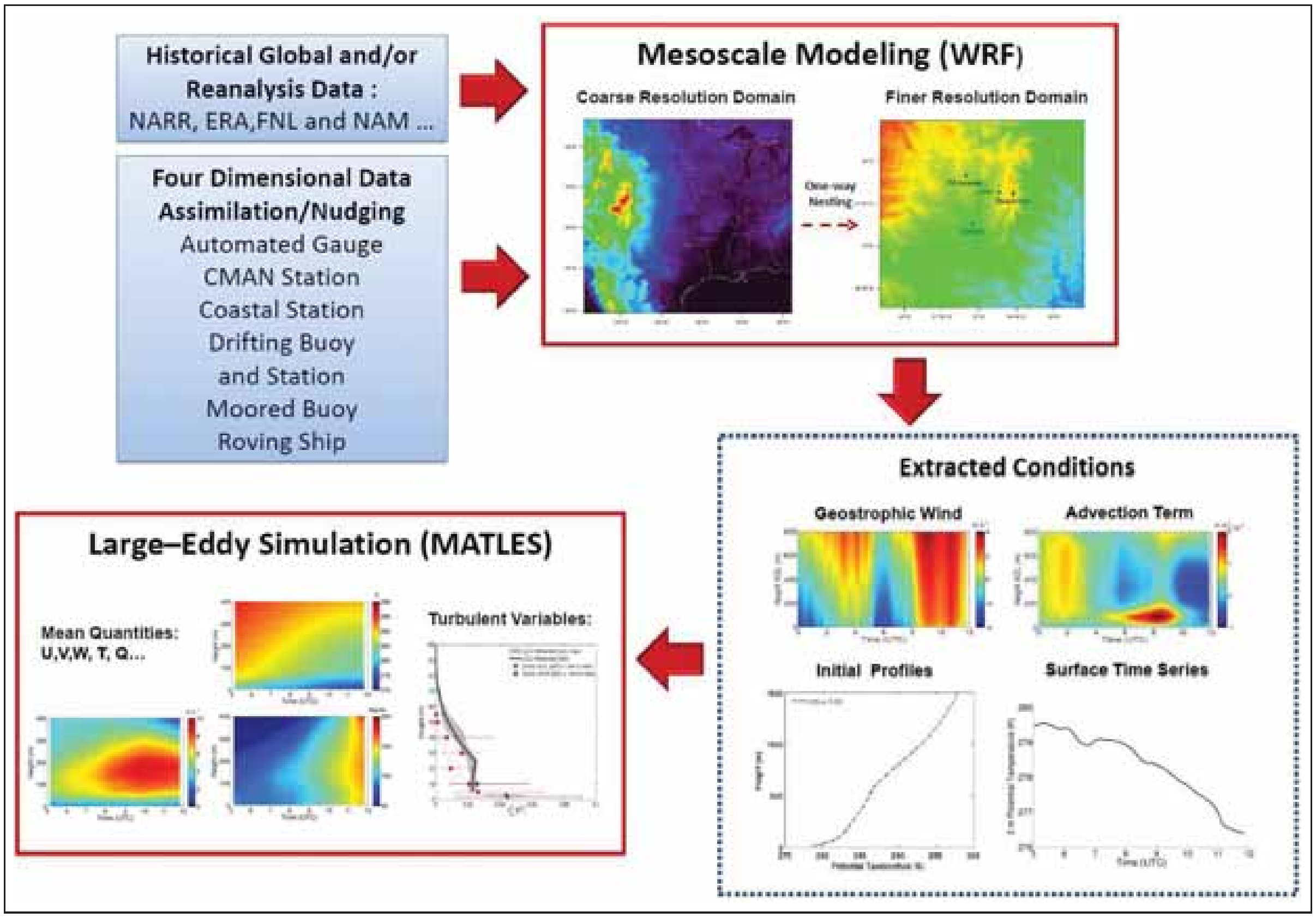

FIG. 1: Procedure for generating realistic turbulence utilizing a coupled mesoscale and large-eddy modeling framework.

.

Rizza et al.[74] did not incorporate radiation physics in their LES model. Other technical differences are in numerical configurations, LES subgrid-scale (SGS) modeling, geostrophic wind estimation strategy, etc. Also, our selected nighttime case study is more scientifically challenging than the daytime case simulated by Rizza et al.[74].

## IV. DESCRIPTION OF CASE STUDY

In this paper, we simulated a nocturnal stable boundary layer observed during the period, October 23-24, 1999 as part of the Cooperative Atmosphere-Surface Exchange Study - 1999 (CASES-99) field campaign[71]. This particular case is quite interesting due to the existence of intermittent turbulence, as well as due to the presence of a moderately strong low-level jet. On this night, the mid-west United States (encompassing the CASES-99 site) was dominated by a high-pressure system (see Fig. 2). At the CASES-99 site, the sky was virtually cloud-free, setting the stage for a moderately/strongly stratified boundary layer. Other synoptic conditions for this case were discussed by Shin and Hong[77] and will not be repeated here for brevity.

The CASES-99 field site was located near Leon, Kansas (37.65° N, 96.74° W). The relatively flat terrain and uniform roughness ($z_\circ = 0.03$ m) of this site make it attractive for boundary layer studies. Please refer to Fig. 3 for a depiction of the topographical features around this field site.

The selected case study was earlier simulated by several researchers utilizing single-column, mesoscale, and large-eddy models: Steeneveld et al.[81], Steeneveld et al.[82], Kumar et al.[46], Shin and Hong[77], and Svensson et al.[89]. Based on these past studies, we were able to better understand the challenges associated with the selected case study. Furthermore, some of these studies helped us to decide on certain mesoscale model configurations (see Section VI for details).

## V. DESCRIPTION OF OBSERVATIONAL DATASETS

In this work, a diverse suite of observational datasets was utilized for model validation. These datasets were collected by a doppler lidar, a small-aperture scintillometer, several sodars, sonic anemometers, and a sounding system. Used in a complementary fashion, these datasets increased the reliability of the nocturnal turbulence characterization. The basic characteristics of these datasets are provided in Table I. For detailed descriptions (including working principles of the instruments, data acquisition, calibration, accuracy, etc.) please refer to the references listed in Table I.

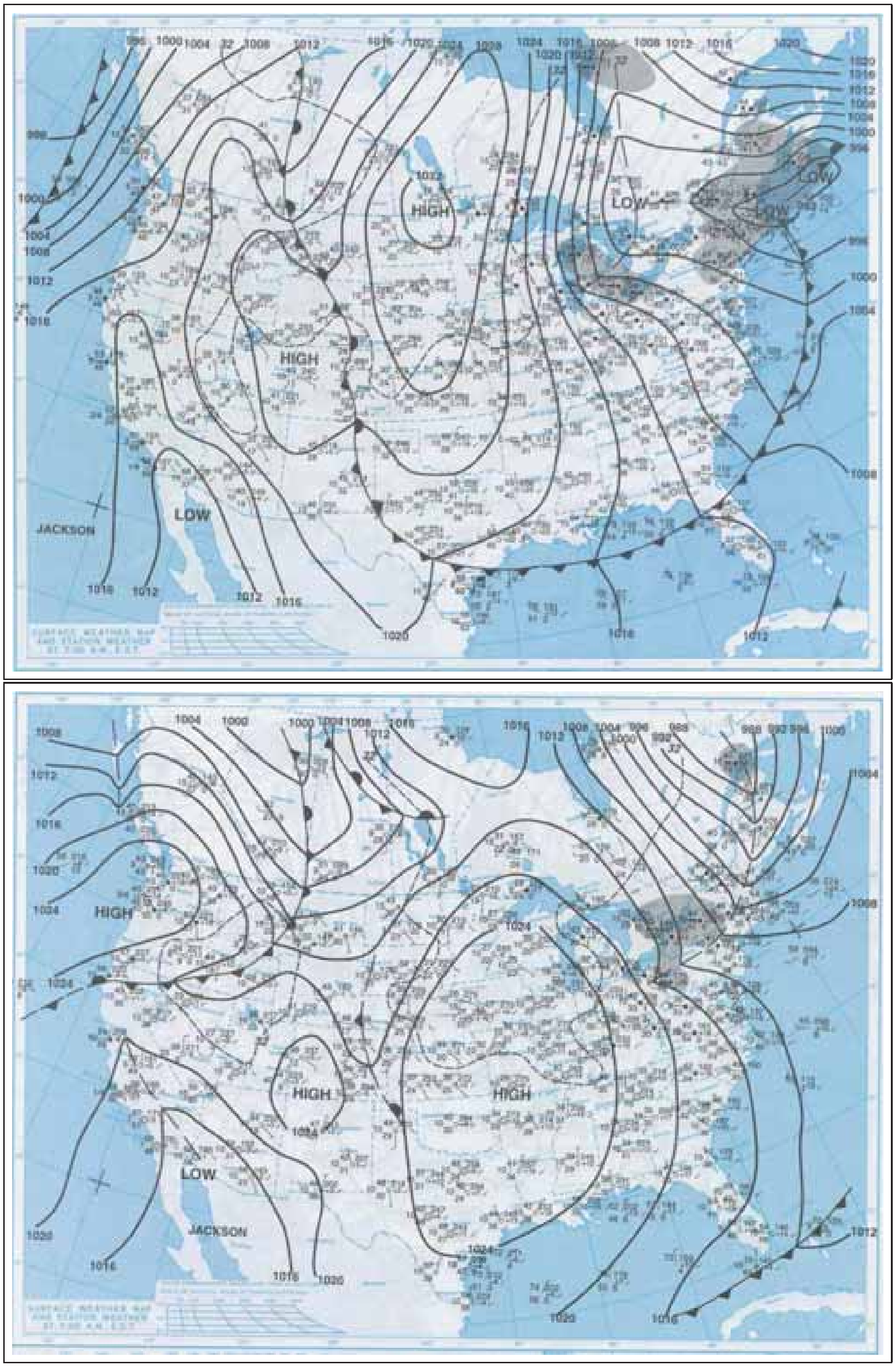

FIG. 2: Surface analysis at 12 UTC on October 23, 1999 (top panel) and October 24, 1999 (bottom panel). The Midwest United States was dominated by a synoptic-scale surface high-pressure system during this time period. Source: http://www.hpc.ncep.noaa.gov/dailywxmap/

TABLE I: Basic characteristics of the observational datasets

| Instrument | Variables* | Frequency | Vertical Coverage | References |
|---|---|---|---|---|
| Lidar | $M$, $\sigma_u^2$ | 1 Hz | $\approx 0-400$ m AGL | [6,69,71] |
| Scintillometer | $u_*$, $\langle w'\theta' \rangle$ | every 6 s | $\approx 2.5$ m AGL | [31,32] |
| Sodar | $M$, $X$ | every 30 min | $\approx 0-200$ m AGL | [18,28,71] |
| Sonic-EOL | $M$, $X$, $\sigma_u^2$, $\sigma_v^2$, $\sigma_w^2$, $\sigma_\theta^2$, $u_*$, $\langle w'\theta' \rangle$ | 20 Hz | $\approx 0-60$ m AGL | [71,88] |
| Sonic-WUR | $M$, $X$, $\sigma_u^2$, $\sigma_v^2$, $\sigma_w^2$, $\sigma_\theta^2$, $u_*$, $\langle w'\theta' \rangle$ | 20 Hz | $\approx 3$ m, 10 m AGL | [31,71] |
| Sounding | $M$, $X$, $\Theta$, $Q$ | a few times per day | several km AGL | [6,71] |

*Relevant variables are defined as follows:

$M$: wind speed (m s$^{-1}$)

$X$: wind direction (degrees)

$\Theta$: potential temperature (K)

$Q$: specific humidity (kg kg$^{-1}$)

$u_*$: friction velocity (m s$^{-1}$)

$\langle w'\theta' \rangle$: sensible heat flux (K m s$^{-1}$)

$\sigma_u^2$: variance of longitudinal velocity component (m$^2$ s$^{-2}$)

$\sigma_v^2$: variance of lateral velocity component (m$^2$ s$^{-2}$)

$\sigma_w^2$: variance of vertical velocity component (m$^2$ s$^{-2}$)

$\sigma_\theta^2$: variance of potential temperature (K$^2$)

The acronyms EOL and WUR in Table I stand for Earth Observing Laboratory of the National Center for Atmospheric Research (NCAR) and Wageningen University and Research Centre, respectively.

## VI. MESOSCALE MODELING

In this study, Version 3.3.2 of the Weather Research and Forecasting (WRF) model was utilized for mesoscale simulations. This state-of-the-art, non-hydrostatic model includes numerous atmospheric physics parameterizations and advanced data assimilation modules[79]. Over the past few years, the WRF model has been developed by the collaborative efforts of multiple organizations such as the National Center for Atmospheric Research (NCAR), the National Oceanic and Atmospheric Administration (NOAA), universities, and others. Several recent studies (e.g.,[24,27,42,64,85]) have demonstrated the strengths of the WRF model in simulating various mesoscale atmospheric phenomena (including LLJs).

In Section IV, we mentioned that Steeneveld et al.[82] performed an intercomparison study of several MMMs for the selected case study. They reported:

> "All schemes underestimate the diurnal temperature cycle amplitude and the near-surface stability at night. None of the parameterizations was able to represent the surface radiation and turbulent fluxes, the wind speed and temperature profiles, and the boundary layer height correctly during the full diurnal cycle. Schemes with local mixing provide a more realistic representation of the nighttime boundary layer, especially for weak winds, and when the asymptotic length scale is based on the flow properties. Moreover, the nighttime low-level jet is hard to reproduce, and we find a clear dependence on the chosen model domain size."[82]

These findings provided valuable guidance in our selection of the WRF domain size and the planetary boundary layer (PBL) scheme. The selection of other numerical settings and physical parameterizations was largely based on our past experience[64,85,86].

We used a large outer domain (resolution: 27 km) of dimensions 2700 km $\times$ 2700 km centered on Leon, KS (see the top left panel of Fig. 3). This domain size was sufficient for reliable LLJ simulations. Three one-way nested domains with resolutions of 9 km, 3 km, and 1 km were set up inside this outer domain. Also, 51 non-uniformly spaced vertical grid levels with approximately 7 levels below 200 m were used (top right panel of Fig. 3).

Shin and Hong[77] conducted an extensive PBL scheme sensitivity experiment for the selected case study. They reported tremendous variability among simulations utilizing different PBL schemes; more importantly, they were unable to identify a clear 'winner' among the various PBL schemes. For this reason, following Occam's razor principle, we chose one of the simplest PBL schemes available with the WRF model—the Yonsei University (YSU) scheme[35,36]. This first-order scheme utilizes the K-profile approach[19,65,84,91] and is numer-ically very stable[16]. Recently, H. Richardson and S. Basu discovered a numerical bug in the YSU scheme. This bug has been corrected in Version 3.4.1 of the WRF model. In the present study, we used an older version of the WRF model in conjunction with the bug fix.

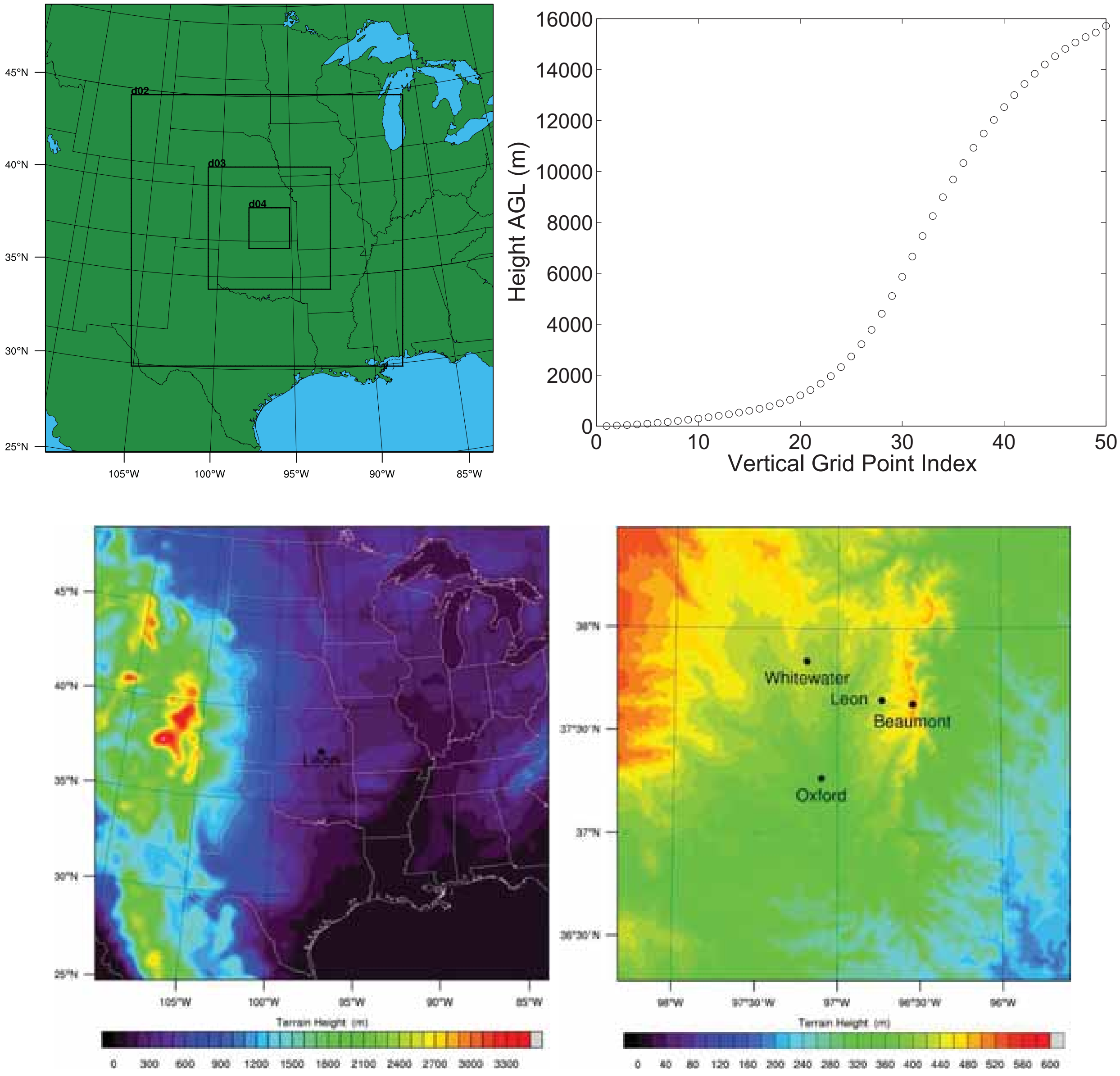

FIG. 3: The WRF model runs utilize four nested domains (top left panel) of grid resolutions 27/9/3/1 km. The locations of the vertical grid points are shown in the top right panel. The elevation maps of the coarsest and the finest domains are shown in the bottom left and bottom right panels, respectively. The locations of various observational sites are marked on these panels.

The following other physical parameterizations were selected in this study: (i) microphysics: WRF Single-Moment 5-class scheme[37]; (ii) shortwave and longwave radiation: Rapid Radiative Transfer Model for Global Climate Models (RRTMG) scheme[38,58]; (iii) cumulus: Kain-Fritsch scheme[40,41]; and (iv) land surface: Noah scheme[23]. The cumulus scheme was switched on only for the domains with 27 km and 9 km resolutions. For the finer resolution domains, the cloud processes were simulated explicitly. These parameterization schemes are discussed in great detail by Stensrud[84].

We performed two mesoscale simulations using the aforementioned numerical and physical settings. In one of the simulations (called WRF-SN), we invoked the so-called four-dimensional data assimilation (FDDA) technique[50,51,80] to assimilate the public-domain NCEP ADP Global Surface Observational Weather Data[1]. We wish to note that the diverse observational datasets described in the previous section were not assimilated; they were solely used for model validation. The FDDA configuration details are summarized below:

- the nudging coefficient for both wind and temperature was set equal to $6 \times 10^{-4}$ s$^{-1}$; this is the default value in the WRF model;
- the horizontal radii of the nudging influence for all the variables were chosen as ten times the grid size (e.g., the horizontal influence of the finest domain is $10 \times 1$ km = 10 km);
- the vertical radii of the nudging influence were defined on the $\eta$ levels and were set equal to 0.002;
- the half-period time width was defined as 10 min (i.e., the WRF model searches for observed data 10 min before and 10 min after the current integration time).

The other mesoscale simulation (called WRF-NN) did not include any data assimilation.

The initial and boundary conditions from the North American Regional Reanalysis (NARR; spatial resolution: 32 km; temporal resolution: 3 h) dataset were used for both the WRF simulations. The simulations started at 12 UTC on 23 October and continued for almost 24 hours; the last 7 h of the simulations (from local midnight to sunrise) are reported on in this paper. The WRF model output was stored every 10 min for a comprehensive analysis and for synthesis of the input data for the LES runs.

[1] http://rda.ucar.edu/datasets/ds464.0/

Most of the modeling results are presented in Section VIII. However, in the present section, we discuss two issues related to the mesoscale modeling of SBLs and LLJs: (i) spatio-temporal error and (ii) spatial variability.

In Fig. 4, the WRF model-simulated wind speeds are compared with lidar-based observations at Leon. The following discrepancies between observed and modeled data are evident:

- the WRF model simulated a weaker and slightly higher LLJ;
- the timings of the observed and modeled LLJ peaks are quite different.

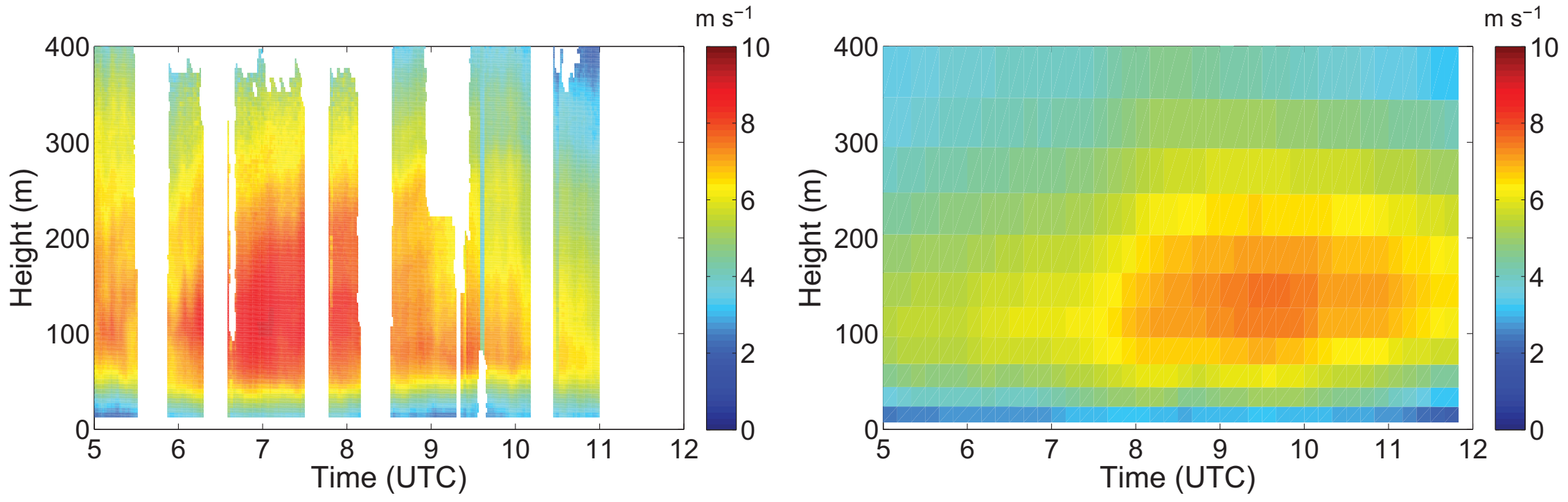

FIG. 4: Time-height plots of lidar-based (left panel) and WRF model-simulated (right panel) wind speeds.

It is well-known from the literature (e.g.,[25,86]) that the MMMs usually create weaker and higher LLJs due to enhanced diffusion. The simulations reported by Steeneveld et al.[82] and Shin and Hong[77] also support this fact. Therefore, the first discrepancy is in line with the literature.

The second disagreement between data and model is, however, unexpected. According to Blackadar's inertial oscillation hypothesis[87,93], the peak of the LLJ at Leon should occur $\approx 9.8$ h after sunset[2]. On October 24, 1999, sunset at Leon occurred at 23:40 UTC; thus, the

[2] The latitude at Leon is 37.65° N. The corresponding Coriolis parameter, $f$, is $8.9 \times 10^{-5}$ s$^{-1}$. The LLJ peak should occur at $\pi/f = 9.8$ h after frictional decoupling at sunset.

LLJ peak maximum was expected to occur around 9:30 UTC. The WRF model simulated the LLJ peak maximum exactly at this time. Interestingly, the observed LLJ peak occurred 3 h earlier. It is likely that some other dynamical mechanisms were responsible for this behavior, which were not captured by the WRF model.

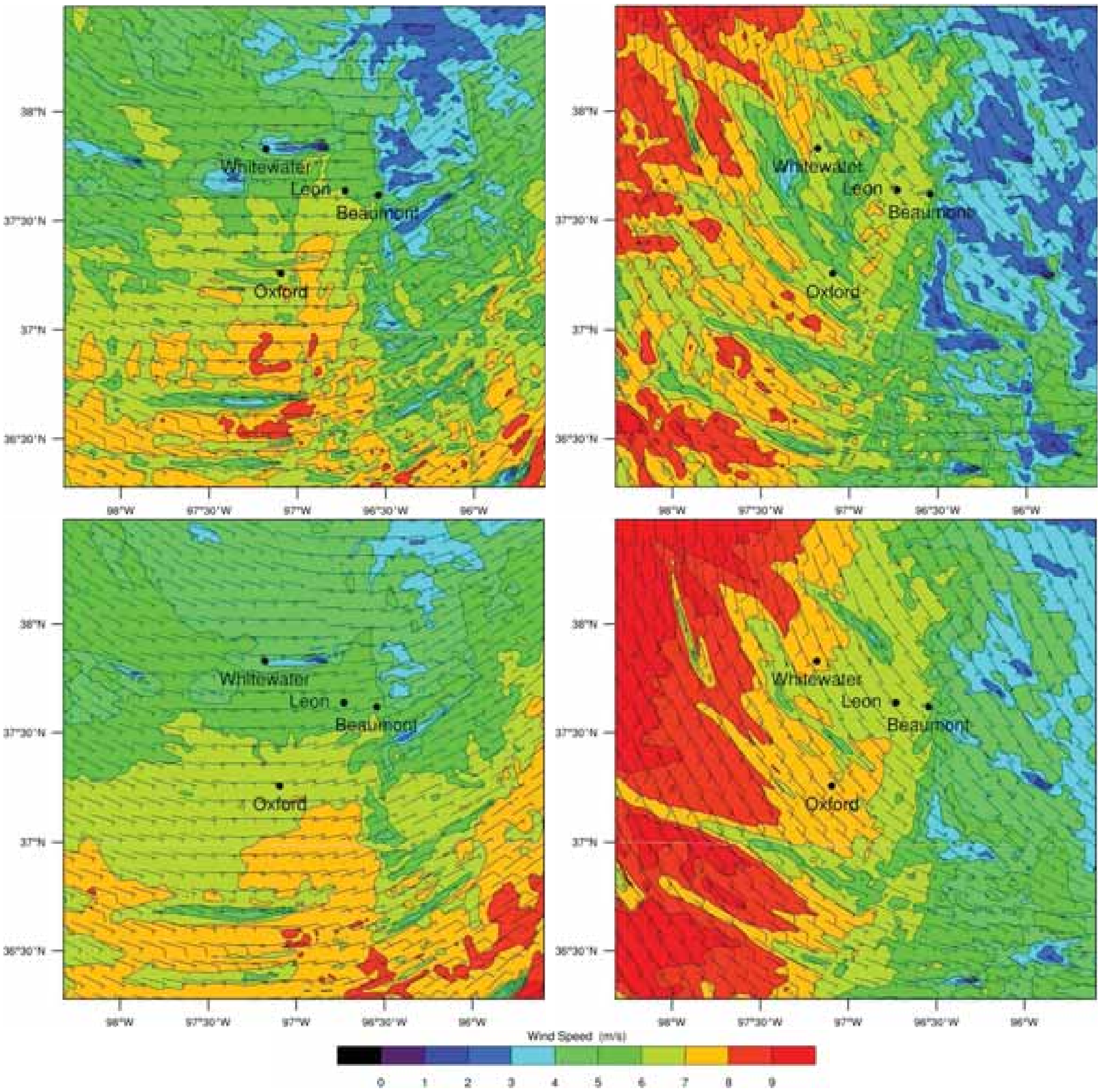

FIG. 5: The WRF model-based simulated wind speeds and wind barbs at 90 m (top panels) and 150 m (bottom panels) above ground level. The left and right panels represent 7 UTC and 11 UTC, respectively, on October 24, 1999. The locations of various observational sites are marked on these panels.

The WRF model-based simulated wind speeds and wind barbs at 90 m (top panels) and 150 m (bottom panels) above ground level (AGL) are shown in Fig. 5. The left and right panels represent 7 UTC and 11 UTC, respectively, on October 24, 1999. From this figure, it is clear that the nocturnal wind fields are extremely variable. Similar conclusions can also be drawn from Fig. 6 which shows sodar-based wind field evolutions from four neighboring locations (on relatively flat terrain): Beaumont, Leon, Oxford, and Whitewater.

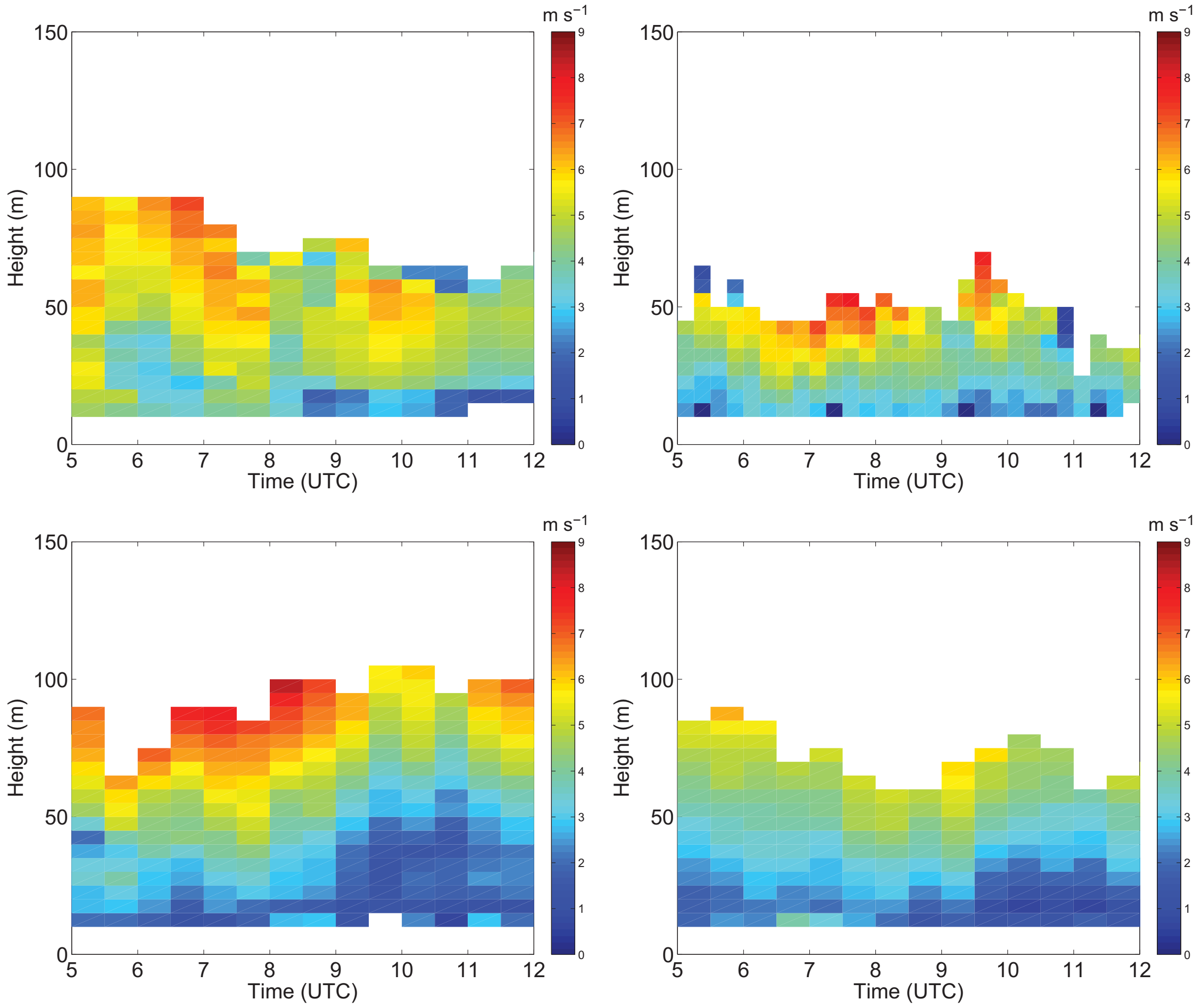

FIG. 6: Time-height plots of sodar-measured wind speeds at Beaumont (top left panel), Leon (top right panel), Oxford (bottom left panel), and Whitewater (bottom right panel).

The spatio-temporal variabilities of wind fields shown in Figs. 4-6 have strong implications from a model validation perspective. We speculate that similar (or perhaps stronger) spatio-temporal variabilities exist in the nocturnal turbulence fields. At the same time, high-quality turbulence observations are only available from a single location: Leon. This lack of spatio-temporal observations causes reliable model validation to be a challenging if not impossible task. Thus, the model validation results presented in Section VIII should be studied with some degree of caution.

## VII. LARGE-EDDY SIMULATION

In this study, an in-house model, MATLES, is utilized for the large-eddy simulations. It utilizes the locally-averaged scale-dependent dynamic (LASDD) subgrid-scale (SGS) model. The most recent version of this model includes Version 2.1.7 of the Column Radiation scheme (CRM;[44]). Other technical details of the MATLES model have been described in various publications (e.g.,[12,73]) and will not be repeated here for brevity.

The LES runs were initialized with the WRF model-generated mean profiles for Leon at 5 UTC (local midnight), October 24, 1999 (not shown). This particular initialization time was chosen to reduce the impact of the residual layer turbulence on the SBL simulations. A similar strategy was used for the GEWEX Atmospheric Boundary Layer Study (GABLS) third LES intercomparison project[13]. In idealized simulations, small-scale random perturbations (noise) are typically added to the initial profiles. In this work, more realistic noise profiles were generated utilizing Nieuwstadt's local scaling approach[10,62]. Specifically, for horizontal velocity fields, we used Gaussian noise with zero mean and variance equal to $4u_*^2\,(1-z/h)^{3/2}$ for $z < h$. However, for the potential temperature field, we used Gaussian noise with zero mean and variance equals to $9\theta_*^2$ for $z < h$. Based on the WRF model-generated profiles of wind speed and potential temperature, the height, $h$, of the SBL at 5 UTC was estimated to be equal to 100 m. The values of surface friction velocity ($u_*$) and surface temperature scale ($\theta_*$) were also extracted from the WRF model-based simulations.

All the simulations were run for $\approx$ 7 h until 11:50 UTC (sunrise time was 12:44 UTC). The lower boundary conditions were based on the Monin-Obukhov similarity theory with a surface roughness length, $z_\circ = 0.03$ m[89]. The WRF model-simulated time series for 2 m potential temperature and 2 m specific humidity were used for lower boundary conditions (see Fig. 7). The overall cooling rate at 2 m was $\approx 0.25$ K h$^{-1}$. The specific humidity remained almost constant over the entire simulation period. During the simulations, the surface sensible heat flux ($\langle w'\theta' \rangle$) was estimated using both the prescribed potential temperature $(\Theta_2)$ and the mean potential temperature $(\Theta(z_1))$ at the model's lowest level $(z_1)$ as follows:

$$\langle w'\theta' \rangle = \frac{\kappa u_* \left[\Theta_2 - \Theta(z_1)\right]}{\log\left(\frac{z_1}{2}\right) - \psi_H\left(\frac{z_1}{L}\right) + \psi_H\left(\frac{2}{L}\right)} \tag{1}$$

where $L$ is the Obukhov length; $\kappa$ is the von Kármán constant (= 0.4); and $\psi_H\left(\frac{z_1}{L}\right) = -5z/L$ for stably stratified conditions[4]. A similar approach was followed for the surface latent heat flux estimation. Please refer to Basu et al.[12] for a description of the surface shear stress estimation procedure.

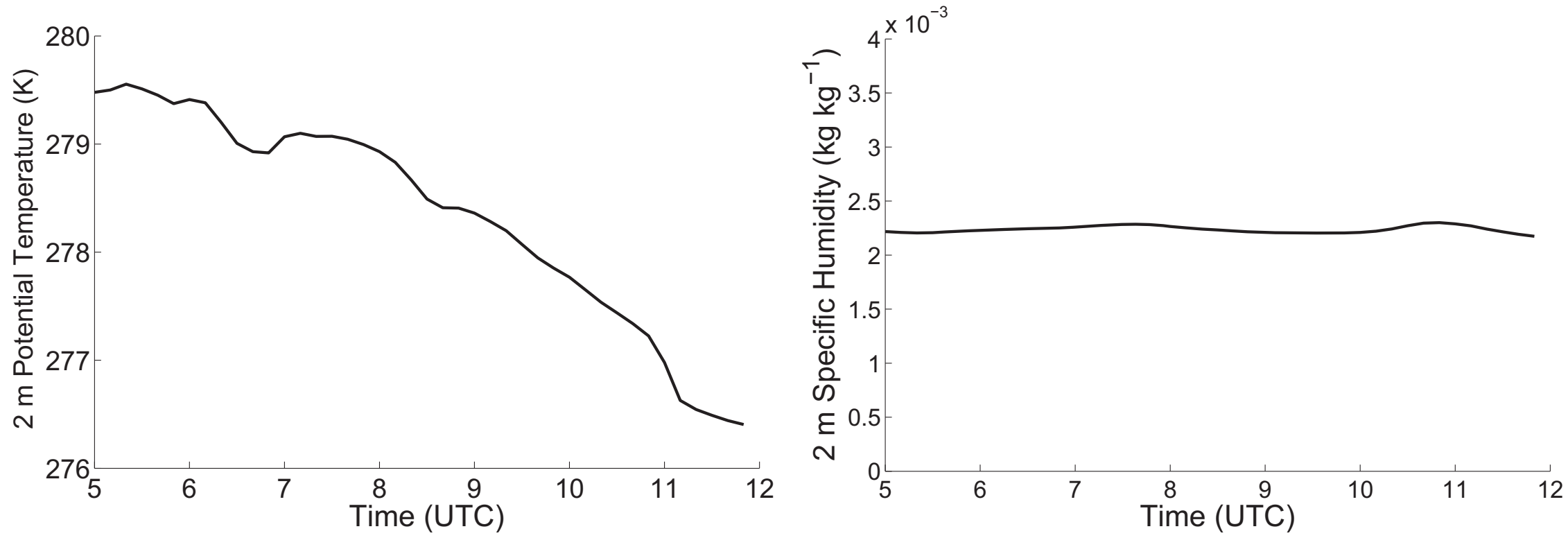

FIG. 7: The WRF model-simulated time series of 2 m potential temperature (left panel) and 2 m specific humidity (right panel). These data were used as lower boundary conditions for the LES runs.

In the past, numerous SBL modeling studies that have been undertaken used the sensible heat flux as a lower boundary condition[21,39,75]. In Basu et al.[11], the fundamental shortcomings of such sensible heat flux-based lower boundary conditions were discussed. Based on analytical and numerical results, it was shown that, if the surface sensible heat flux is prescribed as a boundary condition, only the near-neutral to weakly stable regimes are captured. In order to represent moderate to very stable regimes in simulations, surface temperature prescription or prediction is required. Holtslag et al.[34] provide further insights into this topic. For these reasons, we used near-surface potential temperature as a lower boundary condition.

In contrast to the aerodynamic roughness length $(z_\circ)$, the thermal roughness length $(z_{\circ T})$ is not as well understood and there is no consensus in the literature regarding the $z_\circ - z_{\circ T}$ relationship. Thus, we used 2 m potential temperature instead of the surface temperature; based on Eq. 1, it is clear that $z_{\circ T}$ was not needed in our approach.

The upper boundary consisted of a zero stress condition, whereas the lateral boundary conditions assumed periodicity. A Rayleigh damping layer at 550 m was used. Potential temperature and specific humidity gradients were guided by the WRF output and prescribed at the upper boundary as follows: $d\Theta/dz = 0.0142$ K m$^{-1}$, $dQ/dz = 2.468{\times}10^{-6}$ Kg Kg$^{-1}$ m$^{-1}$.

Time-height-dependent geostrophic wind components $(U_g, V_g)$ were estimated from the WRF model-generated pressure fields (Fig. 8). In order to avoid local-scale (smaller than the meso-$\beta$ scale of 20 km) perturbations, the estimated geostrophic wind fields were spatially filtered (over horizontal planes) using a moving average filter of 20 km $\times$ 20 km stencil size. The presence of strong baroclinicity (likely due to the sloping terrain of the U.S. Great Plains) is evident in Fig. 8 and should always be accounted for in realistic simulations.

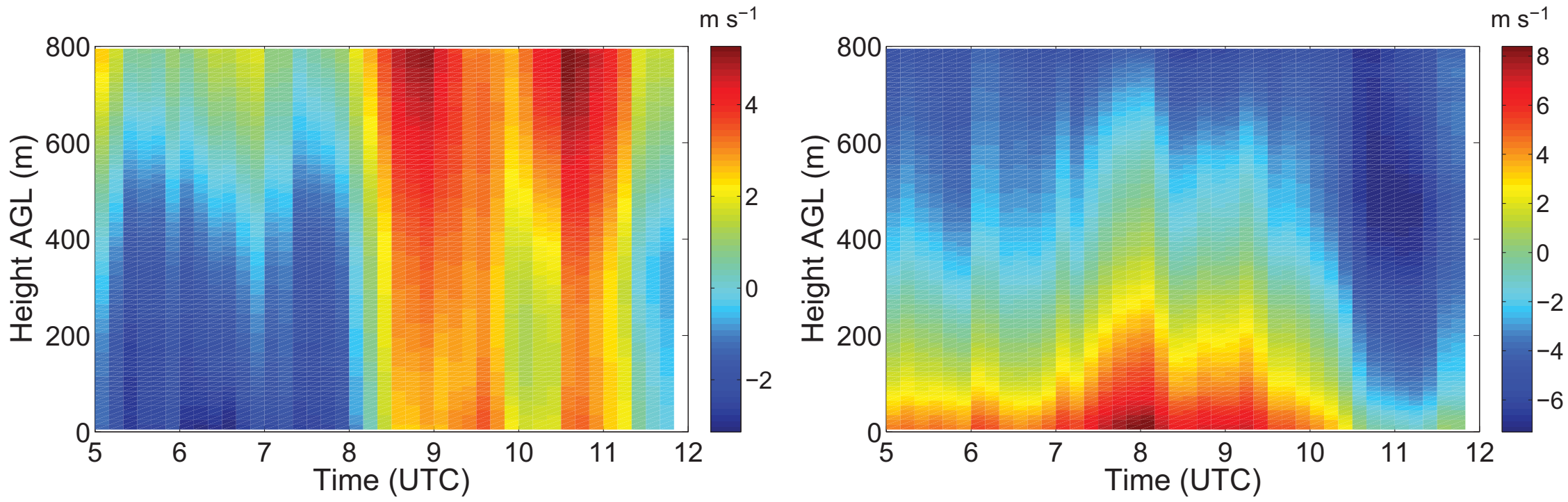

FIG. 8: Time-height plots of the WRF model-simulated zonal (east-west; left panel) and meridional (north-south; right panel) geostrophic wind components.

Similar to the geostrophic wind fields, the time-height-dependent mesoscale advection terms (for momentum, heat, and moisture) were also obtained from the WRF-generated fields (Fig. 9). These fields were also spatially filtered with a stencil size of 20 km $\times$ 20 km. We would like to point out that we neglected the vertical component of advection in our calculations. In other words, for a generic variable, $\Phi$, the advection term is defined as:

$$ADV_\Phi = -\left(U\frac{\partial\Phi}{\partial x} + V\frac{\partial\Phi}{\partial y}\right) \tag{2}$$

Mesoscale advection terms are usually neglected in LES studies (one of the exceptions being the GABLS third LES intercomparison case;[13]). However, these terms could be sig-nificant for realistic simulations. For example, from Fig. 9, it is clear that a large amount of low-level temperature advection (up to $\approx 0.7$ K h$^{-1}$) occurred during 5-12 UTC. This additional heat source reduced the bulk stability and generated a deeper SBL. The following section discusses this topic in greater detail.

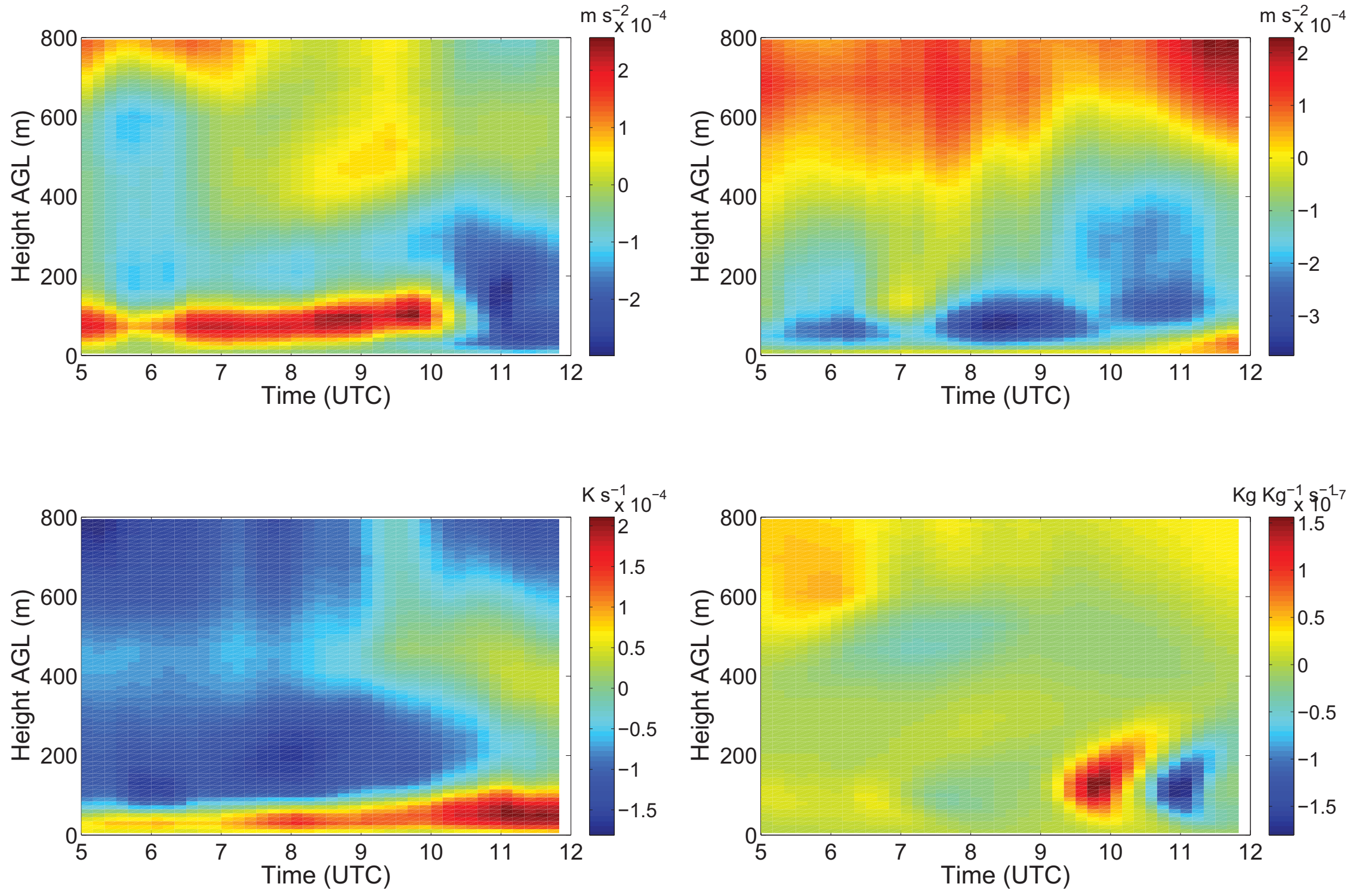

FIG. 9: Time-height plots of the WRF model-simulated mesoscale advection terms: zonal wind component (top left panel), meridional wind component (top right panel), potential temperature (bottom left panel), and specific humidity (bottom right panel).

Recently, Mirocha and Kosović[57], studied the influence of subsidence on SBL simulations. The impacts of changing the subsidence from zero to 0.002 m s$^{-1}$ was substantial in terms of the depth, mixing, and cooling rate of the SBL. Svensson et al.[89] also considered a constant subsidence of 0.005 m s$^{-1}$ in their simulation of the GABLS second single-column model intercomparison study. They estimated subsidence from a mesoscale output. In the present work, we also attempted to estimate subsidence from the WRF model output (see Fig. 10). Subsidence was found to be spatio-temporally highly intermittent. Moreover, we were unable to isolate the topographical effect from the synoptic-scale atmospheric effect.

Given the uncertainty in its estimation, we decided not to include subsidence in the LES runs. In retrospect, we believe that this exclusion resulted in slightly deeper SBLs in the LES runs in comparison to the WRF model results.

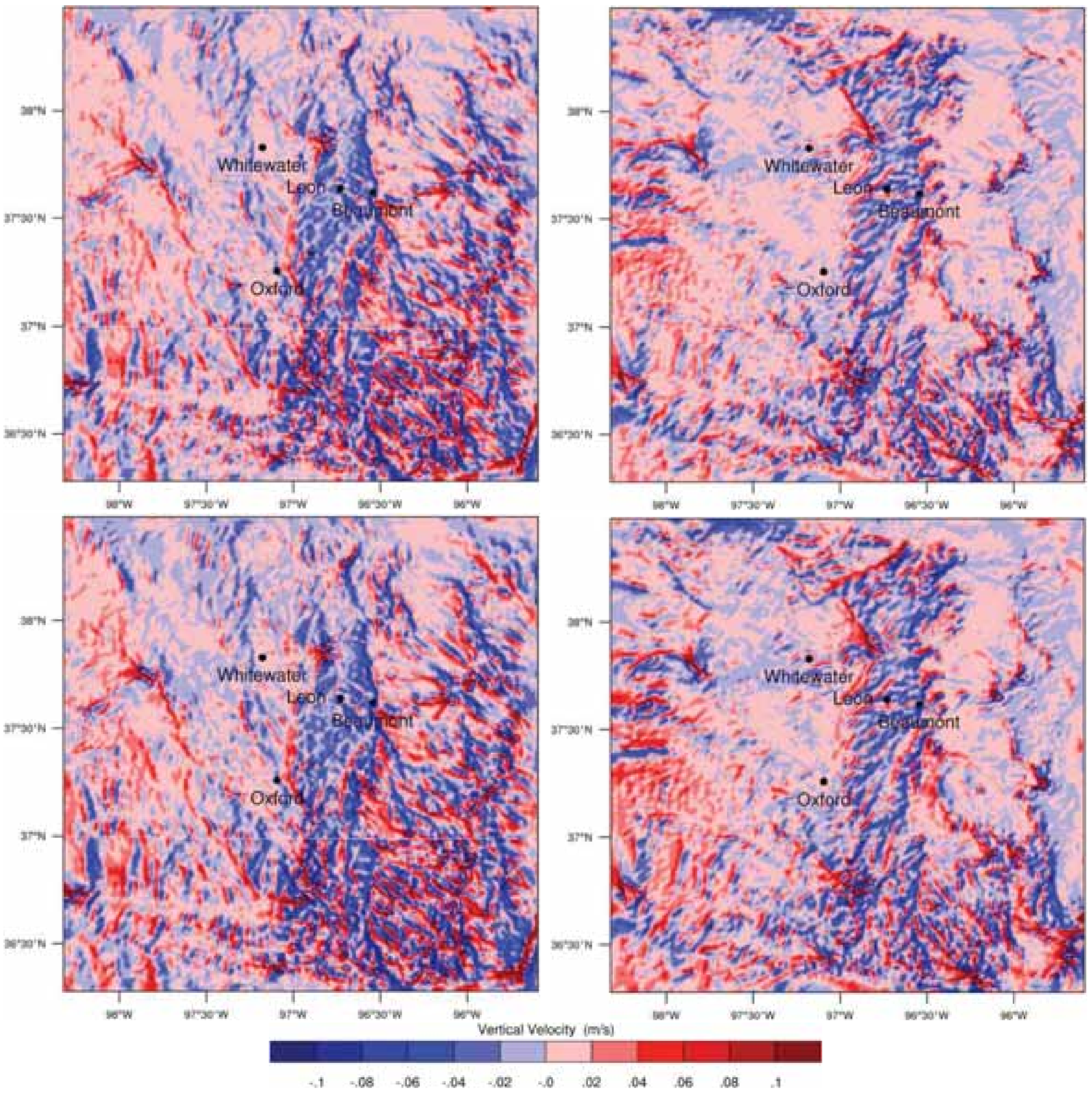

FIG. 10: The WRF model-based simulated vertical velocity at 90 m (top panels) and 150 m (bottom panels) above ground level. The left and right panels represent 7 UTC and 11 UTC, respectively, on October 24, 1999. The locations of various observational sites are marked on these panels.

The selected domain size for all the LES runs was 800 m × 800 m × 790 m, divided into 80 × 80 × 80 grid points (i.e., the grid resolution was 10 m × 10 m × 10 m). The land surface was assumed to be flat which is a realistic assumption for the selected location near Leon. The time step was equal to 0.125 s, while each run was ≈7 h (i.e., 196,800 time-steps) in duration. In order to assess the influence of the grid resolution on the results, we performed two runs with 64 × 64 × 64 grid points ($\Delta x$ = 12.5 m × $\Delta y$ = 12.5 m × $\Delta z$ = 12.5 m; $\Delta t$ = 0.2 s) and 40 × 40 × 40 grid points ($\Delta x$ = 20 m × $\Delta y$ = 20 m × $\Delta z$ = 20.3 m; $\Delta t$ = 0.4 s), respectively. In addition, two more LES runs were performed to assess the impacts of radiation and mesoscale advection on the simulated results. Table II summarizes the configurations of all the LES runs. Planar-averaged profiles were output every 10 min from these simulations. For spectral analysis, time series data from selected grid points were output every time step.

TABLE II: List of LES runs and associated configurations

| Run | Spatial Resolution | Radiation Scheme | Mesoscale Advection |
|---|---|---|---|
| L80-A-R-SN | 10.0 m×10.0 m×10.0 m | Yes | Yes |
| L80-A-SN | 10.0 m×10.0 m×10.0 m | No | Yes |
| L80-SN | 10.0 m×10.0 m×10.0 m | No | No |
| L64-A-R-SN | 12.5 m×12.5 m×12.5 m | Yes | Yes |
| L40-A-R-SN | 20.0 m×20.0 m×20.3 m | Yes | Yes |

## VIII. RESULTS

In this section, simulated mean profiles from the WRF and MATLES models are first compared. The similarity of the results would indicate that the forcing terms (i.e., geostrophic winds and mesoscale advection terms) were appropriately extracted from the WRF output. It would also indicate that the physical parameterizations (e.g., microphysics), which are present in the WRF model, but absent in the MATLES model, were not important for this clear-sky case.

Next, the modeled results are validated against a diverse set of observations. Given the spatio-temporal variabilities of the observations and the simulated results (see Section VI), as well as the apparent inconsistencies among different observational platforms (discussed below), we argue that it is not necessary to place undue emphasis on the quantitative differences. On the other hand, qualitative trends and some similarity statistics (e.g., in temporal spectra) should provide valuable insights.

## A. First-Order Statistics

The time-height plots of the simulated wind speed, wind direction, potential temperature, and specific humidity are shown in Fig. 11. For the WRF model results, the vertical profiles represent instantaneous values from the grid point closest to Leon. In the case of the MATLES results, these profiles represent planar-averaged values from the entire domain.

The timing of the LLJ event as well as the temporal evolution of the wind direction profiles are similar in both the mesoscale and the LES results. This qualitative similarity suggests that the dynamical forcings are similar in both the simulations. However, the LLJ is stronger and slightly higher in the case of the LES results. It is widely known that the strength of the LLJs is strongly dependent on the diffusion of the associated SBL. Given that the strength of the LES-generated jet is more similar to the observed one (see the left panel of Fig. 4) than is the case for the WRF-generated one, we conclude that the LES has captured the nighttime mixing with greater fidelity than the YSU PBL scheme of the WRF model. Since the LES run did not include subsidence, the simulated LLJ height is slightly higher. This feature is also visible in the time-height plot of the potential temperature—the LES generated a slightly deeper SBL. The evolution of specific humidity is very similar in the mesoscale and LES results. Due to accurate prescription of the mesoscale advection, the LES run has even captured a short burst of moisture enhancement during 10-11 UTC. This enhancement could be due to a small-scale frontal passage.

As the observed sounding data were available at 7 and 11 UTC, we chose those specific times for comparison of the observed and simulated vertical profiles (Figs. 12 and 13). In addition to the sounding data, we overlaid data collected by a doppler lidar (wind speed only) and a meteorological tower (wind speed and wind direction). It is quite interesting to see that the lidar-based wind speed observations are quite different in magnitude from the sounding observations. Banta et al.[6] mentioned some technical issues related to the sounding launches during the CASES-99 field campaign, which might explain the differences. Since other publications (e.g.,[77,82]) compared simulated results against these soundings, we decided not to exclude them in this paper.

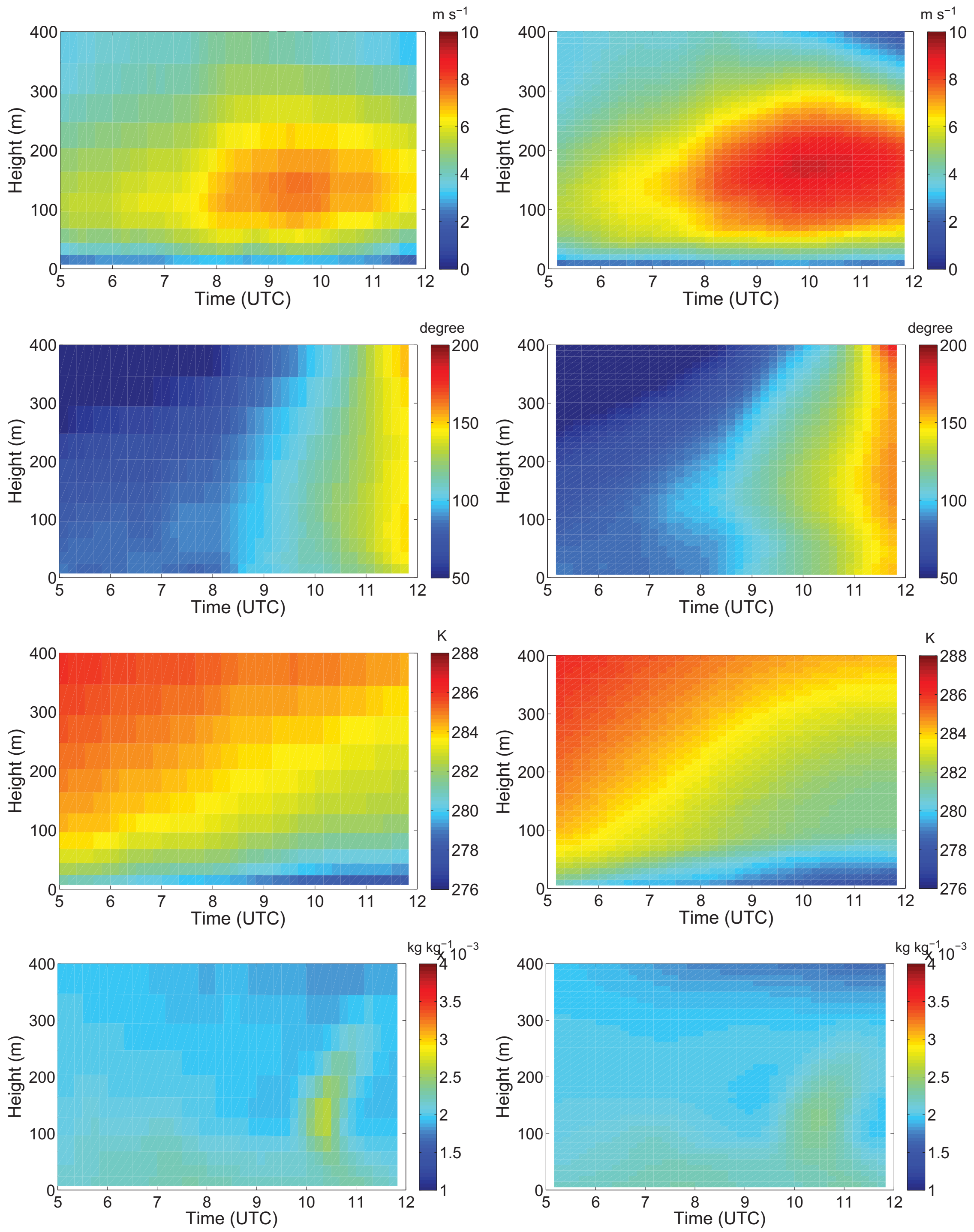

FIG. 11: Time-height plots of simulated wind speed (top panels), wind direction (second panels), potential temperature (third panels), and specific humidity (bottom panels). The left and right panels represent results from the WRF model and the MATLES model, respectively.

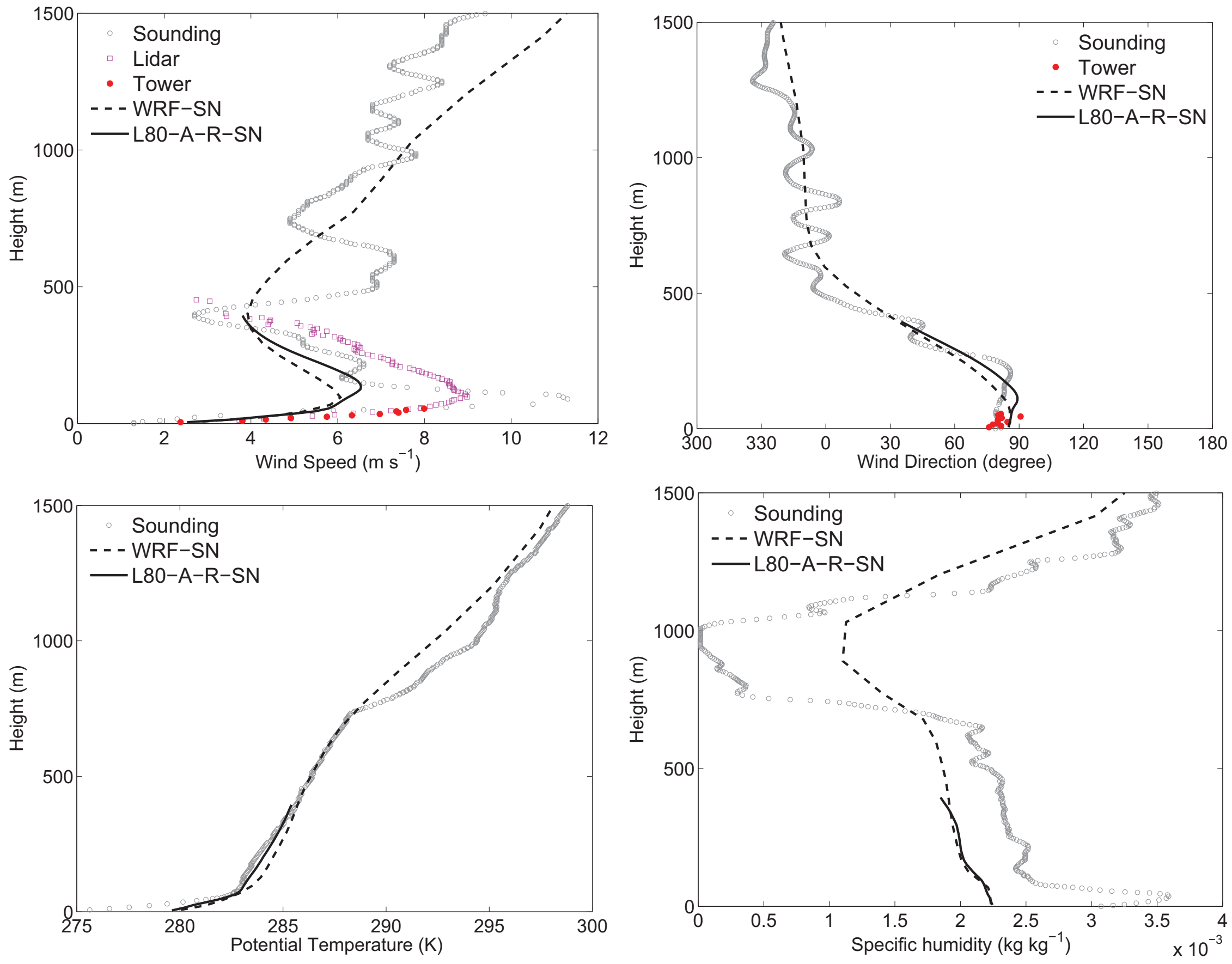

FIG. 12: Plots of the wind speed (top left panel), wind direction (top right panel), potential temperature (bottom left panel), and specific humidity (bottom right panel) profiles corresponding to 7 UTC.

Considering, first, the profiles from 7 UTC presented in Fig. 12, we note that the heights of the lidar-based and the WRF-based LLJs are quite similar. As explained before, the LES-based LLJ is slightly higher due to the lack of subsidence. The observed and simulated wind shears agree very well in the lower part of the SBL. Differences in the modeled and observed LLJ are largely due to the temporal shift in their evolution.

The WRF and LES models captured the wind direction and potential temperature profiles remarkably well. The wind direction within the SBL ($h \approx 100$ m) was from the east; however, at higher elevations, the wind was more north-north-westerly. The observed and modeled potential temperature profiles portray a three-layered structure. Near the surface, a strong inversion is present in the observed and the modeled data. Above this inversion layer, a weakly stable residual layer is present up to $\approx 750$ m. The free atmosphere with moderately strong stratification is present on top of the residual layer.

The observed specific humidity profile shows an interesting multilayer structure: a near-surface moist layer, a well-mixed residual layer, a drying zone above the residual layer, and a moist-free atmosphere. The WRF model qualitatively captures this multilayer structure; however, there is significant room for improvement. The LES-based profile is indistinguishable from the WRF profile.

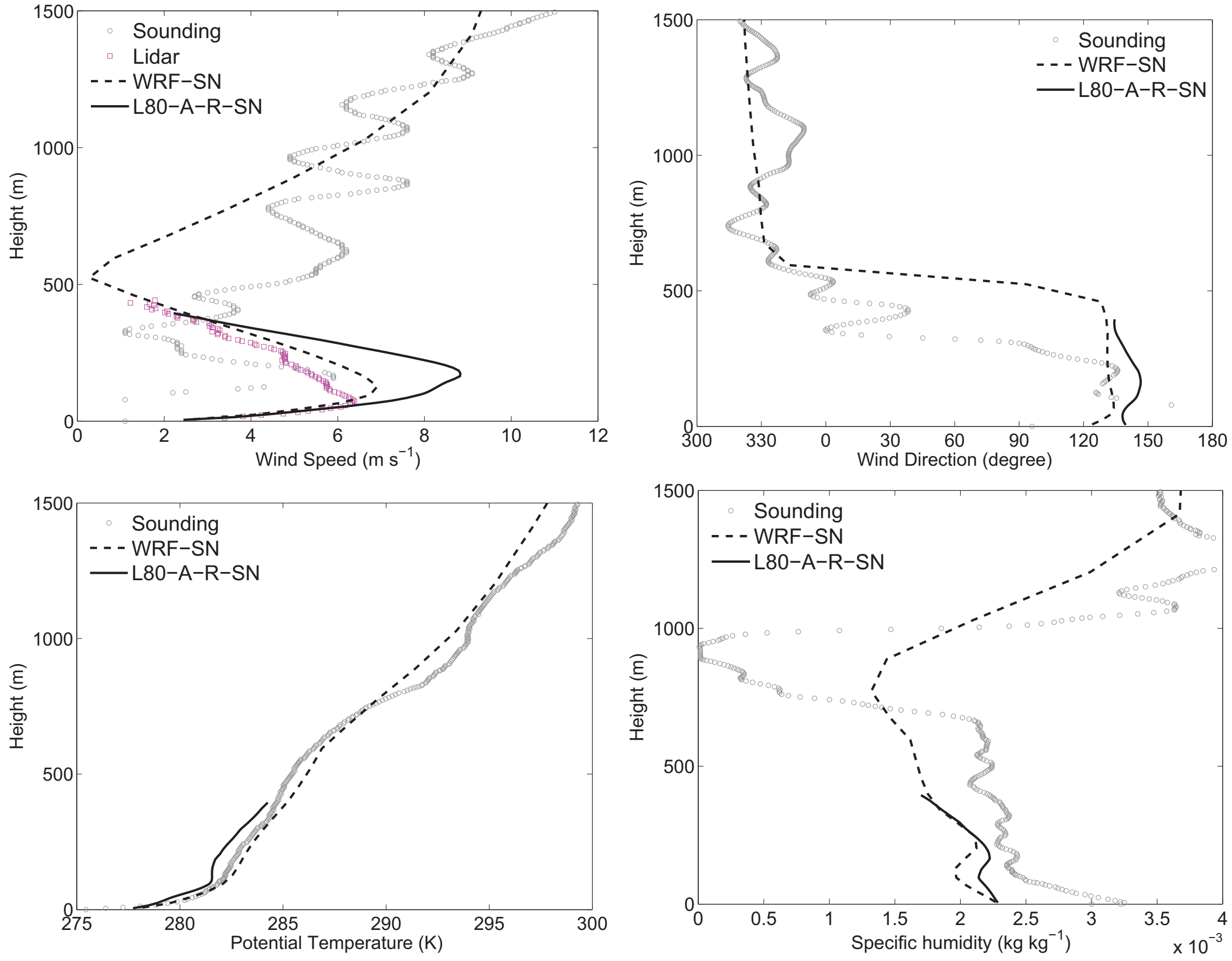

FIG. 13: Plots of the wind speed (top left panel), wind direction (top right panel), potential temperature (bottom left panel), and specific humidity (bottom right panel) profiles corresponding to 11 UTC.

Figure 13 shows profiles from 11 UTC. Almost all the remarks made in the context of Fig. 12 also hold for this figure and, thus, are not repeated. However, we would like to note that the wind direction in the SBL shifted towards south-south-east at this time due to inertial oscillation even though the upper layer wind was still from the north-north-westerly direction.

In Fig. 14, we plot hodographs from observational (0-12 UTC) and modeled (5-12 UTC) data. Since lidar data did not include wind direction information, we utilized sodar and tower wind data. Note that the tower data were not available after ≈ 8 UTC. The signature of inertial oscillations[87,93] is clear in the observed and modeled hodographs. The hodographs from the mesoscale and LES runs are almost identical—emphasizing, again, the similar dynamical evolutions.

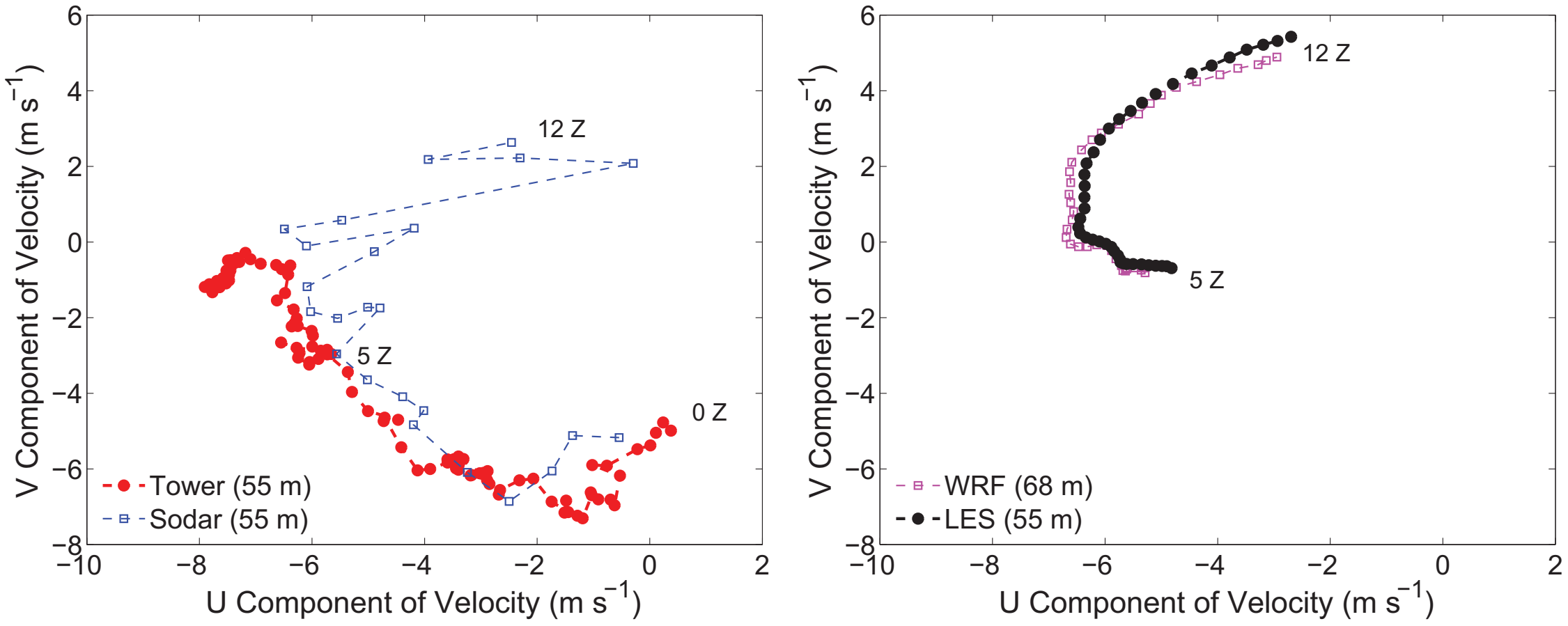

FIG. 14: Observed (left panel) and simulated (right panel) hodographs. In the left panel, the red dots represent velocity observations (with temporal averaging of 5 min) measured by a sonic anemometer located on the 60 m tall meteorological tower (55 m AGL). Wind measurements from a sodar at Beaumont (with temporal averaging of 30 min; 55 m AGL) are plotted as blue squares in the left panel. Simulated data from the WRF model and the MATLES model are represented by magenta dots and black squares, respectively, in the right panel. The signature of inertial oscillations is visible in both the panels.

## B. Second-Order Statistics

Before elaborating on these results, we would like to briefly describe our plotting strategy for variance and flux profiles from the LES runs. As mentioned before, the LES data were saved every 10 min. Thus, for every hour, there are 6 data samples corresponding to each vertical level. From these data points, we calculate and plot the median (p50) values. In addition, the entire range (minimum to maximum) of the LES values is shown as a shaded band. The sonic anemometer data from the EOL and WUR towers are available every 5 min. Analogous to the LES profiles, we also report the median in conjunction with the minimum and maximum values from these datasets.

Observed and simulated variance profiles are shown in Fig. 15. Note that the MATLES model does not solve a prognostic equation of turbulent kinetic energy (TKE); for this reason, we only show the resolved variances from LES. Since EOL tower data are not available after 8 UTC, we report results from two consecutive time frames: 6-7 UTC and 7-8 UTC. We did not report turbulence data from 5-6 UTC as this time was considered to be part of the spin-up period for the LES runs.

From the modeled variance profiles it is quite clear that the nocturnal turbulence is generated near the surface and transported upwards. A non-traditional upside-down character[53,66], where turbulence is generated in the outer SBL rather than at the surface, does not appear in our simulation. In contrast, the horizontal variance observations at 50 and 55 m levels, during 6-7 UTC, might indicate the existence of an upside-down SBL.

There are some differences between the variances computed from measurements by the EOL and WUR sonic anemometers. These could be the result of small-scale topographical effects (the towers were a few tens of meters away from each other) or they could be attributed to differences in instrumentation and/or in the variance calculation.

The observed and modeled horizontal velocity variance and potential temperature variance become very small above $\approx$ 50 m AGL. This height is much shallower than $h_{LLJ} =$ 100 m. In idealized LES studies, $h$ estimated from vertical profiles of different variables (e.g., wind speed, wind-speed profile curvature, stream-wise velocity variance, vertical velocity variance, momentum flux, buoyancy flux) are usually very similar (e.g.,[73]). In contrast, in the case of observational data, different profiles can lead to significantly different estimates of $h$ (see[70,83,96] and the references therein). It was interesting to find out that an LES model with appropriate forcings can also generate different variables with different $h$ values.

The magnitude of the simulated horizontal velocity variance is larger than the corresponding observed value (especially during 7-8 UTC). This could be due to an inherent limitation of the LES approach, or a result of the omnipresent spatial variability. In Fig. 16, we plot streamwise variance from a doppler lidar. Recently, Pichugina et al.[69] reported good correlation between sonic anemometer and lidar data for weakly stable conditions. However, in the present case, the lidar-based variances are much larger than the sonic anemometer-based variances. In other words, there is tremendous discrepancy between the sonic anemometer-and doppler lidar-based estimates of variances; the LES values are somewhat in between the two.

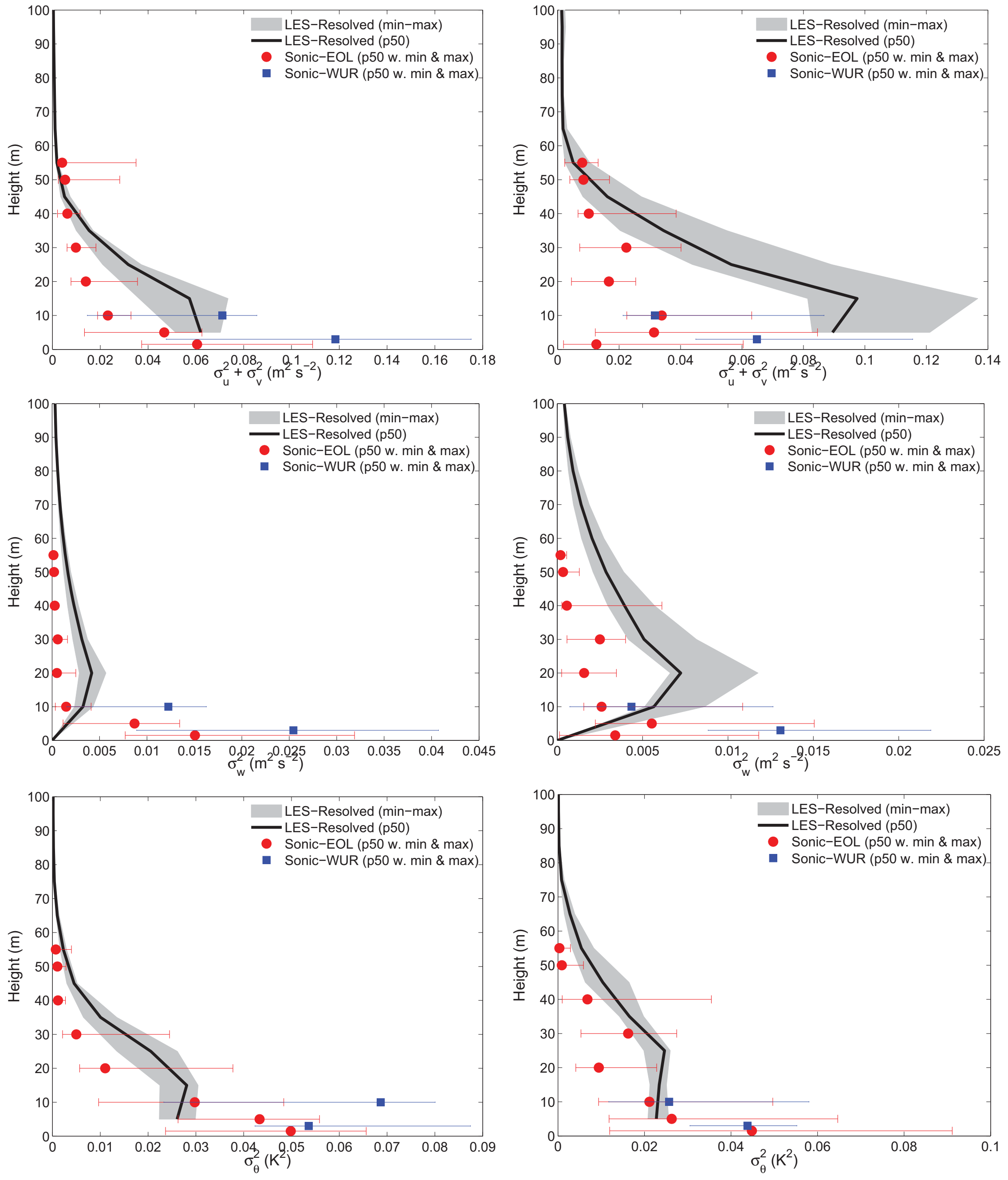

FIG. 15: Plots of variance profiles corresponding to 6–7 UTC (left panels) and 7–8 UTC (right panels). The top, middle, and bottom panels represent horizontal velocity variance ($\sigma_u^2 + \sigma_v^2$), vertical velocity variance ($\sigma_w^2$), and potential temperature variance ($\sigma_\theta^2$), respectively. The red dots with whiskers represent median and minimum-to-maximum values of the observations from the 60 m tall meteorological tower. The solid black lines and the light grey areas correspond to the medians and the minimum-to-maximum ranges, respectively, of the LES-generated output data. Note that the simulated results represent resolved variances, whereas the observed data correspond to total variances.

The LES model underestimates the vertical velocity and temperature variances near the surface. This is likely due to the lack of spatial resolution. We will revisit this issue later in this paper.

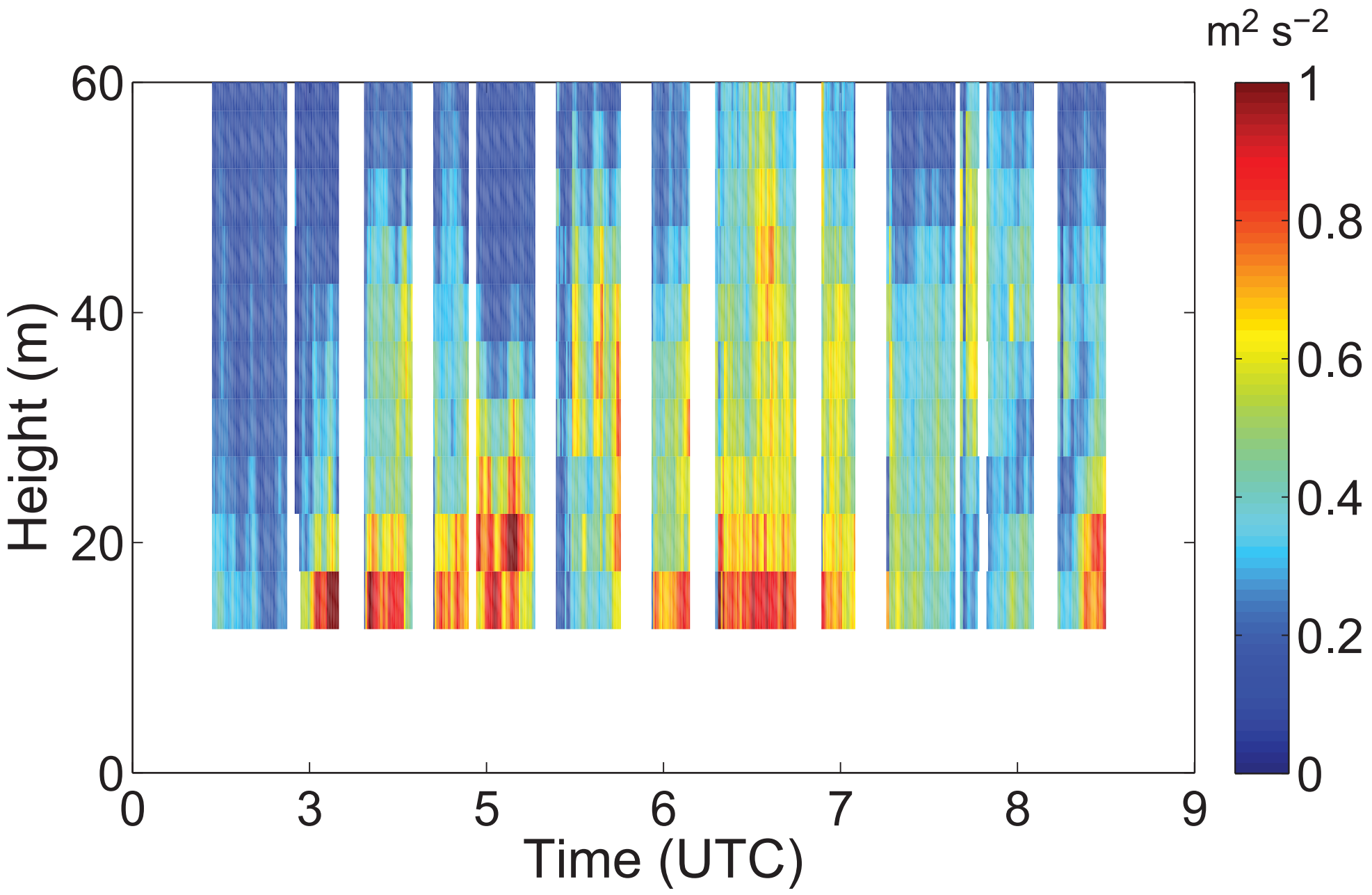

FIG. 16: Time-height plot of doppler lidar-based variances.

Vertical profiles of the local friction velocity ($u_{*L}$), computed using local momentum fluxes, and the sensible heat flux ($\langle w'\theta' \rangle$) are shown in Fig. 17. In contrast to the variance plots, in these plots, the total LES-generated fluxes (resolved plus SGS) are shown. In addition, the resolved fluxes alone are overlaid for comparison.

As with the variance plots, there are some differences between the fluxes measured by the EOL and WUR sonic anemometers. The exact cause for these differences is unknown to us; however, we can speculate and 'blame' the differences on small-scale topographic effects, instrumentation, and/or flux calculations.

Even though the LES captures the near-surface observed fluxes during 6-7 UTC, it overestimates them at higher levels and also during 7-8 UTC. The resolved component of the fluxes is much smaller than that of the corresponding SGS fluxes. For a moderately/strongly stratified case, this behavior is expected from an LES run using a grid resolution of 10 m.

A finer grid resolution will undoubtedly lead to a larger contribution of resolved fluxes; it might also reduce the spurious pile-up of momentum fluxes above 50 m AGL. However, a finer grid resolution is not expected to change the total flux profiles in any substantial manner; we will provide supporting evidence later in this paper.

The modeled variance and flux profiles are consistent with each other—they both reduce to zero or small residual values above $\approx$ 50 m AGL. However, the observed variance and flux profiles behave differently. The behavior of the observed sensible heat flux is especially intriguing—it vanishes almost completely above 20 m AGL. It is, thus, no surprise that Vickers and Mahrt[96] reported $h$ to be barely 20 m at 4 UTC, based on similar buoyancy flux profiles. It is not clear to us whether, by simply refining grid resolution, the LES model will be able to capture this behavior. We will attempt to resolve this issue in future work.

## C. Temporal Evolution

The night of October 23/24 was classified as an intermittent turbulent night by van de Wiel et al.[94]. In this section, we investigate whether the coupled mesoscale-large-eddy modeling approach managed to capture this intermittency in turbulence. We utilized two types of turbulent flux measurements for model validation.

Traditionally, sonic anemometry (also known as the eddy-covariance approach) is used for measuring turbulent fluxes[29]. Recent micrometeorological studies[26,32,55,56] have demonstrated, however, that the use of scintillometry is a viable alternative to sonic anemometry. A scintillometer consists of a transmitter and a receiver and employs the principle of 'scintillation'—turbulence-induced fluctuations of the observed intensity of a remote light source. Over the years, different types of scintillometers have been developed using different wavelengths, aperture sizes, and configurations (see the reviews by[2] and[33] for more information). With a small-aperture scintillometer (SAS), one can estimate the structure parameter of the refractive index, $C_n^2$, and the inner scale of turbulence, $l_0$, from measured amplitude fluctuations. Note that $C_n^2$ and $l_0$ are directly related to the temperature structure parameter, $C_T^2$, and the kinetic energy dissipation rate, $\epsilon$. By using Monin-Obukhov similarity theory, one can extract momentum and sensible heat fluxes from $C_T^2$ and $\epsilon$.

During the CASES-99 field campaign, researchers from Wageningen University (WUR), the Netherlands, deployed a displaced-beam SAS instrument (SLS20 by Scintec). The trans-mitter and receiver were installed at 2.46 m AGL. The path length between the transmitter and the receiver was 112 m. Please refer to Hartogensis et al.[32] for more information on the SAS experimental setup.

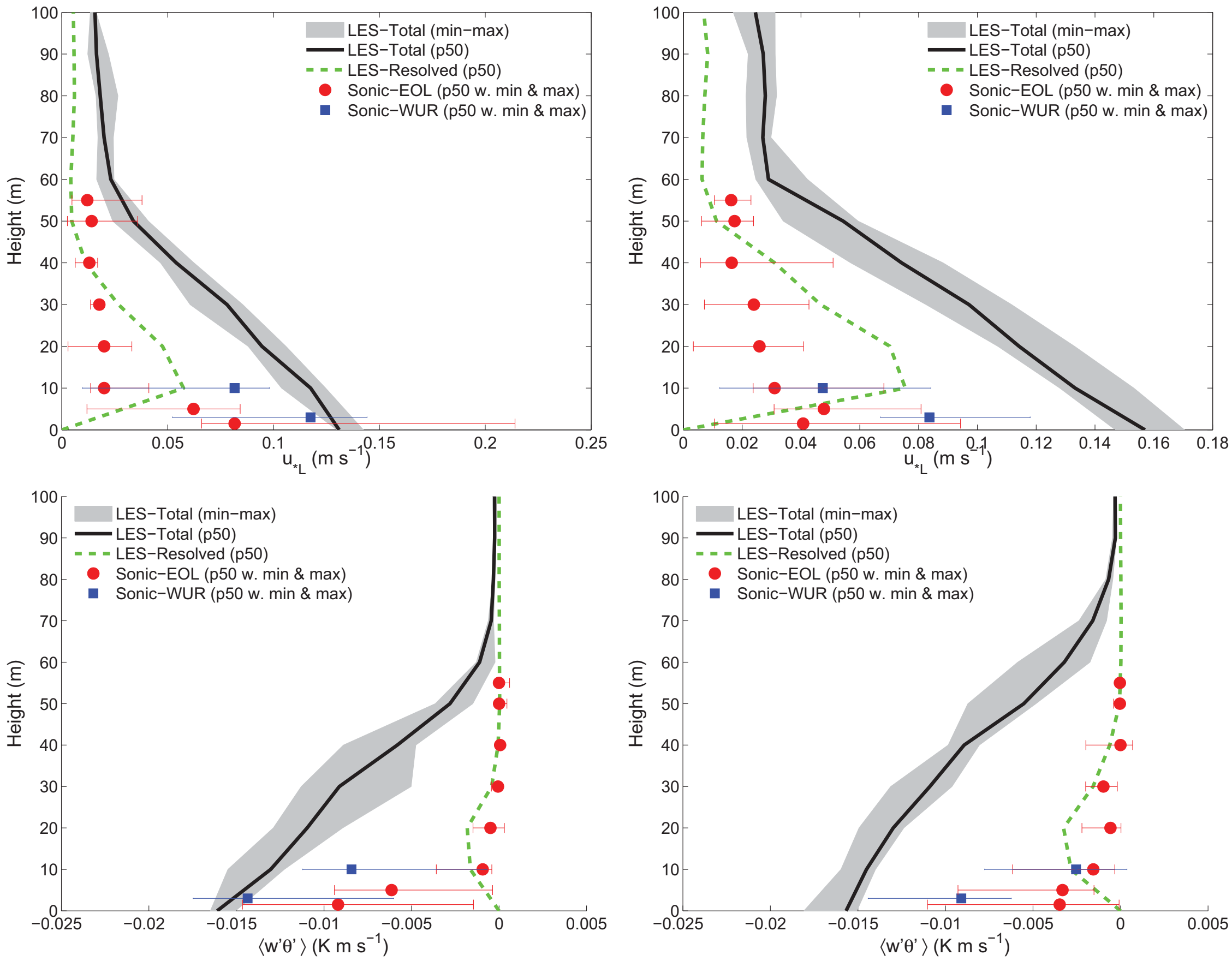

FIG. 17: Plots of local friction velocity (top panels) and sensible heat flux (bottom panels) profiles corresponding to 6–7 UTC (left panels) and 7–8 UTC (right panels). The red dots with whiskers represent median and minimum-to-maximum values of the observations from the 60 m tall meteorological tower. The solid black lines and the light grey areas correspond to the medians and minimum-to-maximum ranges of the LES-generated output data, respectively.

In contrast with conventional sonic anemometers, a SAS requires short averaging intervals (because spatial averaging over a line of sight relaxes the need for long temporal averaging) and is expected to respond rather quickly to changing (non-stationary) atmospheric conditions. During the CASES-99 field campaign, a sampling interval of 6 s was used by the WUR team[32].

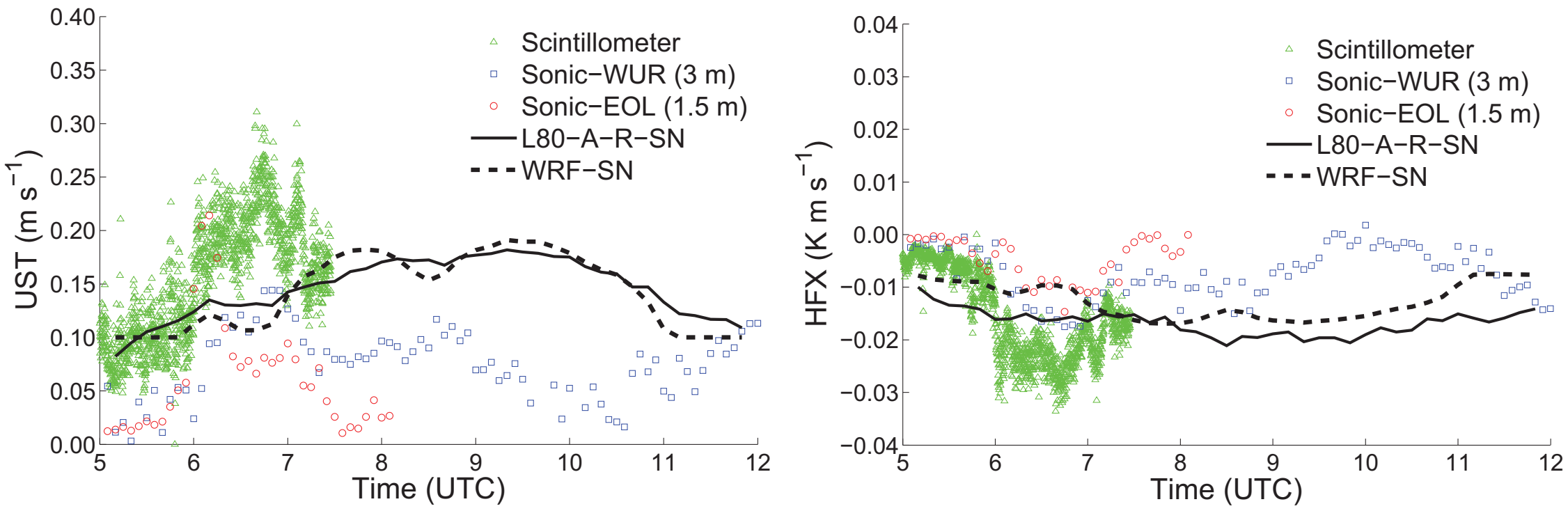

FIG. 18: Time series of surface friction velocity (left panel) and sensible heat flux (right panel). Fluxes measured by a small-aperture scintillometer are depicted with green triangles. Observations from sonic anemometers at 1.5 m and 3 m AGL are represented by stars and circles, respectively. The solid blue and red lines denote the output from the WRF model and the MATLES model, respectively. The observed fluxes clearly portray intermittent behavior.

In Fig. 18, the observed and modeled surface fluxes are shown. Surprisingly, the SAS-based fluxes were significantly larger in magnitude than the sonic anemometer-based fluxes; more interestingly, they were even larger than the modeled fluxes. It is possible that the scintillometer faithfully captured the spatial variability of nighttime surface fluxes. In that case, the SAS-based fluxes (representing path-averaged fluxes) are more realistic than the sonic anemometer-based fluxes (representing point observations). Another possibility is that the Monin-Obukhov similarity functions utilized by the scintillometer are problematic in the intermittent turbulence regime. If that is the case, then the SAS-based fluxes are not trustworthy. More research is definitely needed in this scientific arena.

The observed surface fluxes clearly show signs of intermittency. In contrast, the WRF- and MATLES-generated fluxes are more or less continuous in time. The YSU surface layer scheme of the WRF model artificially clips $u_*$ at 0.1 m $s^{-1}$ to avoid the so-called runaway cooling problem. This spurious behavior is clearly visible in the left panel of Fig. 18. Given that during 6-11 UTC, the WRF-based friction velocity ($u_*$) is much higher than 0.1 m $s^{-1}$, we believe that the clipping has not impacted our results in any significant manner. Nevertheless, in our future work, we will study the impact of this clipping and other thresholding operations on the simulation of intermittent turbulence.

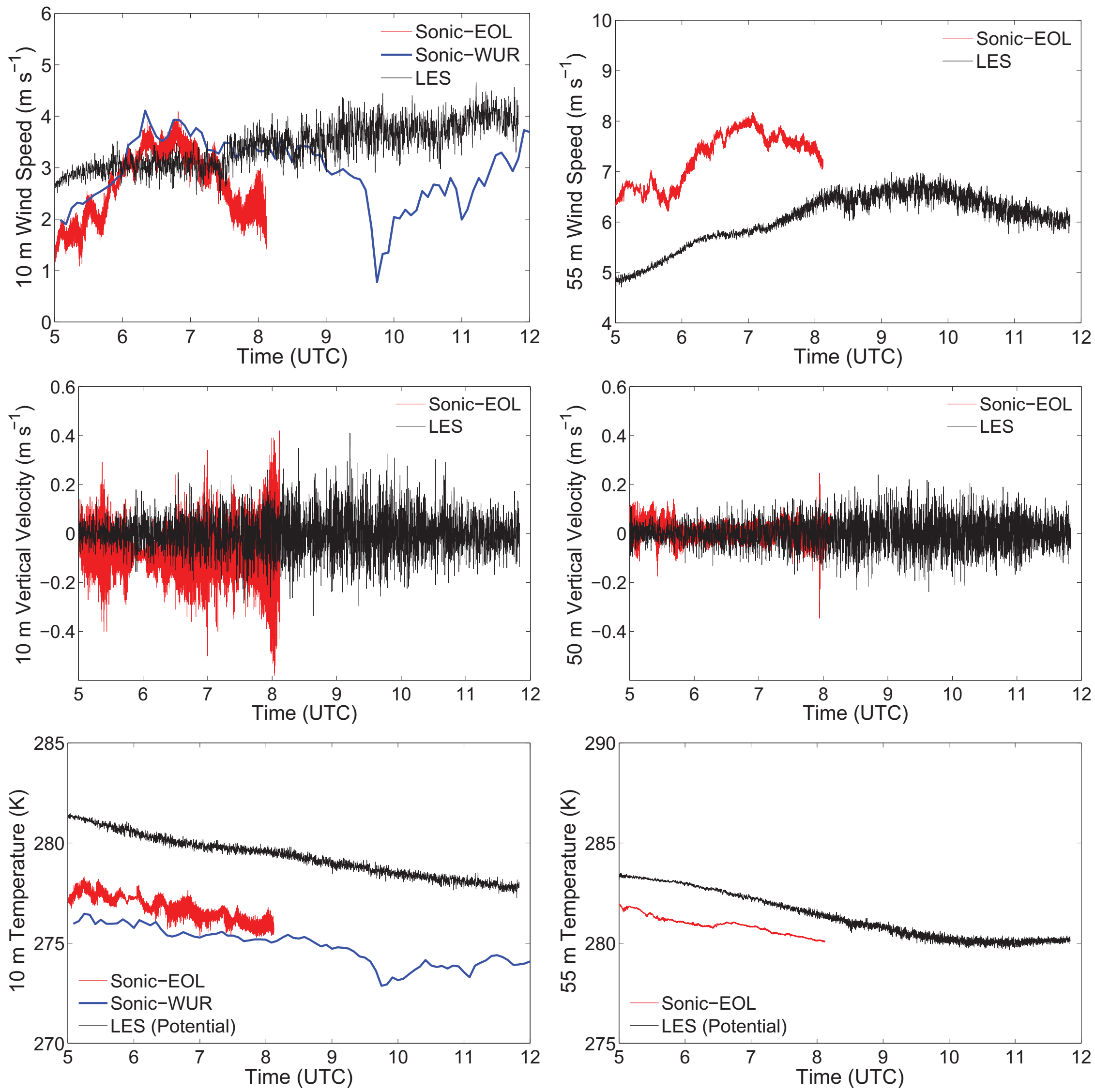

FIG. 19: Time series of wind speed (top panels), vertical velocity (middle panels), and temperature (bottom panels). The left and right panels correspond, respectively, to the near-surface layer (10 m) and the outer layer (55 m for wind speed and temperature; 50 m for vertical velocity).

The WRF-simulated results are in complete agreement with those reported by Shin and Hong[77]. All of their WRF simulations (without exception) utilizing various PBL and SL schemes overestimated nighttime ($u_*$) values when compared against observed sonic anemometer data. At 5 UTC, their simulated $u_*$ values were in the range, 0.15-0.2 m s$^{-1}$. By the end of their simulations (i.e., at 12 UTC), the $u_*$ values decreased to approximately 0.10-0.15 m $s^{-1}$. Similarly, all the WRF simulations also overestimated the magnitude of (downward) sensible heat flux ($\langle w'\theta' \rangle$) values. The simulated $\langle w'\theta' \rangle$ values remained more or less constant during the period of 5-12 UTC. Shin and Hong[77] pointed out that the simulated nighttime surface variables were quite convergent. In fact, the standard deviation for the different simulations was only 0.01 m/s for friction velocity; they were on the order of 4-5 W $m^{-2}$ in the case of sensible heat flux (see Table 2 of[77]).

In Fig. 19, we plot time series of observed and LES-generated horizontal wind speed, vertical velocity, and temperature. In the case of LES, the time series represent simulated data (sampling rate, 10 Hz) from a grid point at the center of the model domain. As noted before, sonic anemometer data from EOL (frequency, 20 Hz) were not available after $\approx$ 8 UTC. Hence, for model validation, we also plotted the 5-minute-averaged data from the WUR sonic anemometer.

For most of the variables, the LES qualitatively captured the range of fluctuations; however, the trends were notably different from the observations. A temporal shift of 3 h is evident in the upper-level (55 m) horizontal wind speed data (see the discussion earlier in the context of inertial oscillation).

Both the observed horizontal wind speed and the vertical velocity display non-stationary character. The LES-generated time series show qualitatively similar behavior; the non-stationary character captured in the simulations is attributed to use of realistic boundary and forcing conditions. For example, soon after the intensification of the simulated LLJ (around 8 UTC), the wind speed fluctuations increased significantly due to shear-generated turbulence (top panels of Fig. 19). The observed vertical velocity time series show some evidence of intermittency. Unfortunately, such intermittency patterns were not present in the LES-generated data.

The near-surface (10 m) observed temperature and horizontal wind speed data show some periodic oscillations. Similar oscillations were lacking in the modeled (potential) temperature and wind speed data. Results reported by Steeneveld et al.[82] and Shin and Hong[77] also did not show any oscillations. According to Revelle[72] and van de Wiel[92], these oscillations are strongly coupled to the phenomenon of surface flux intermittency. Clearly, the mesoscale and large-eddy models are deficient in this regard.

A few minutes before 10 UTC, the WUR sonic anemometer data show signs of the runaway cooling phenomenon. Based on Figs. 18 and 19, we see evidence in the observations of: (i) decoupling of the atmospheric boundary layer from the underlying surface (sharp decrease in the near-surface wind speed; turbulent fluxes become negligible); and (ii) rapid cooling of the near-surface air temperature. The WRF and MATLES models were unable to capture these features.

## D. Spectral Analysis

Traditionally, wavenumber spectra are utilized to assess the strengths and weaknesses of LES-SGS models. It is well documented that non-dynamic (i.e., static) SGS models are over-dissipative as indicated by steeper spectral slopes at higher wavenumbers. On the other hand, in the case of the dynamic SGS models (such as the LASDD SGS model), the longitudinal velocity and scalar spectra clearly show extended inertial ranges (see[1,12] for examples). To the best of our knowledge, the characteristics of LES-generated spectra at larger scales (e.g., the mesoscale, the terra-incognita regime described by[97]) are not discussed in the literature.

A few years ago, Muschinski et al.[61] analyzed observational data from the CASES-99 field campaign. They utilized high-resolution (200 Hz) turbulence data from the CIRES Tethered Lifting System (52–74 m AGL). Their frequency-domain spectra showed three distinct regimes (see bottom right panel of Fig. 20): an inertial range (slope ≈-5/3), a spectral gap (slope ≈0), and a mesoscale range (slope ≈-5/3). It was interesting to assess how our coupled mesoscale-LES approach captured these scaling regimes. We used sonic anemometer data, collected by NCAR-EOL for comparison. Since the observed and modeled time series were non-stationary, we used a discrete wavelet transform (Haar wavelet) approach to compute the frequency-domain spectra. A similar approach was used by Katul et al.[43] and Basu et al.[9]. Results are presented in Fig. 20 and are quite intriguing. The following comments can be made based on this figure:

- The sonic anemometer-based horizontal wind speed and temperature spectra portray slopes of -3 in the mesoscale regime; this is a hallmark of two-dimensional turbulence[48]. In strongly stratified conditions, due to the lack of vertical diffusion, eddies can become quasi-two-dimensional (pancake-shaped); in such cases, one would intuitively expect a -3 slope in the spectra. Since Muschinski et al.[61] analyzed data from a different night,

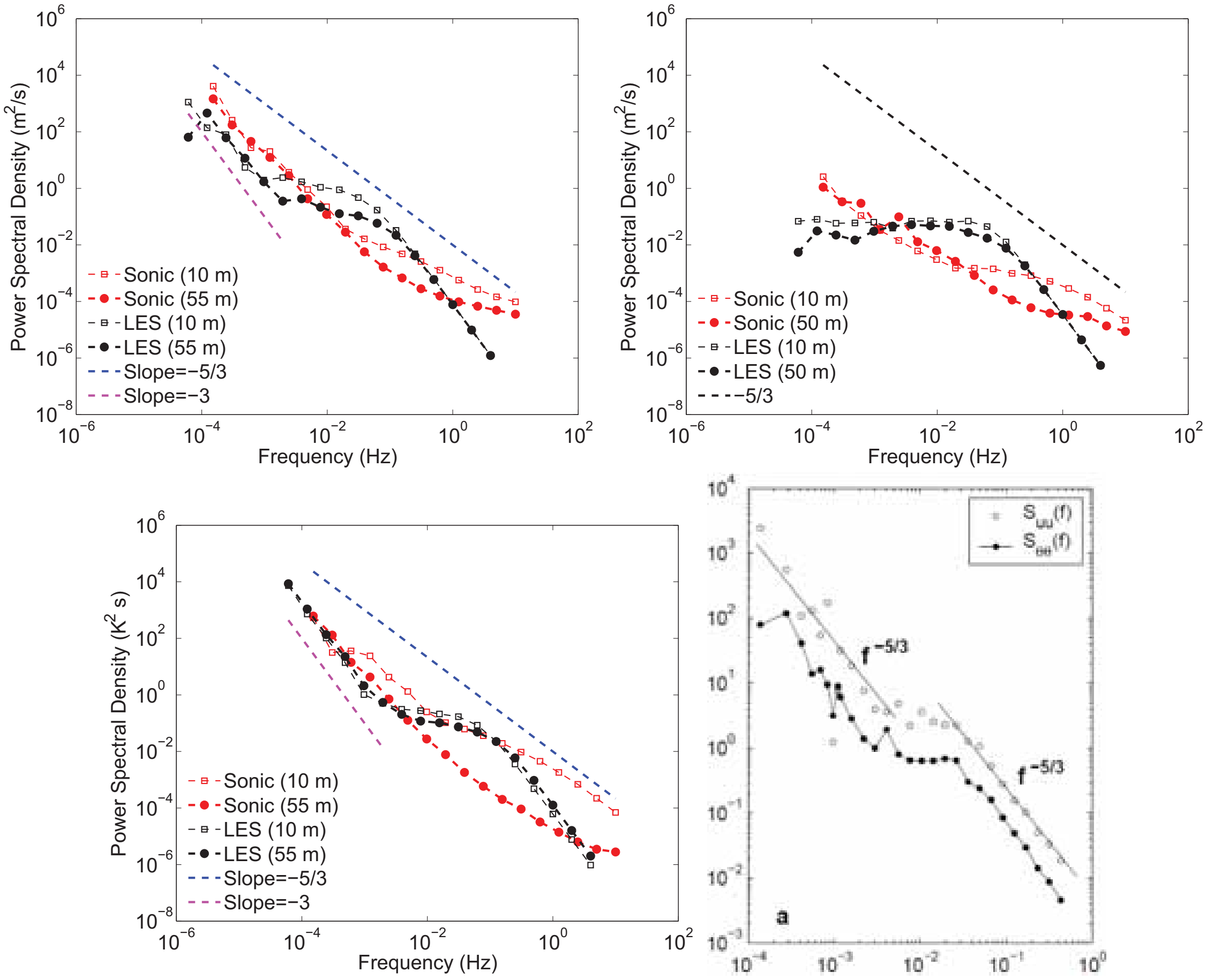

FIG. 20: Wavelet spectra of horizontal wind speed (top left panel), vertical velocity (top right panel) and temperature (bottom left panel). For comparison, spectra reported by Muschinski et al.[61] are shown in the bottom right panel (units are m$^2$ s$^{-2}$ Hz$^{-1}$ for $S_{uu}(f)$ and units, corrected relative to [61], are K$^2$ Hz$^{-1}$ for $S_{\theta\theta}(f)$). Spectral gaps (around 0.01 Hz) are noticeable in the spectra for wind speed and temperature.

it is quite possible that they captured a weakly stable regime with three-dimensional motions.

- In agreement with the observational spectra, in the case of horizontal wind speed and temperature, the modeled mesoscale spectra also show slopes of -3. This specific result significantly boosts our confidence in the simulated data.

- For the high-frequency (inertial) range, most of the sonic anemometer data-based spectra become flat; this is perhaps due to measurement noise. In very stable conditions, sonic anemometers can suffer from several types of instrumental and sampling problems (e.g., dropouts, insufficient amplitude resolution)—please refer to Vickers and Mahrt[95] for further discussion on this topic.

- The LES-generated spectra show steeper than -5/3 slopes in the inertial range. This discrepancy in the inertial range is expected to reduce with increased spatial resolution; however, such simulations would be computationally quite expensive. As a viable alternative, for engineering applications, a fractal interpolation approach could be used to recover the energy in this range (see[9,78] for details).

- The location of the modeled spectral gap (around 0.01 Hz) is identical with the results reported by Muschinski et al.[61]. In the case of the sonic anemometer data, the spectral gaps are somewhat discernible for the near-surface vertical velocity and the temperature data. Note that the existence of a spectral gap around 0.01 Hz was also reported earlier by Caughey[22]. This information was used by several researchers (e.g.,[10,63]) to partition mesoscale and turbulent motions prior to flux estimation.

- The observed vertical velocity spectra appear to follow a -5/3 scaling. In contrast, the modeled vertical velocity spectra display a flattening behavior in the low-frequency range. We speculate that, by increasing the model domain size, we might improve the shape of these spectra.

## E. Sensitivity Experiments

In this section, we report results from various sensitivity experiments.

### *1. Observational Data Assimilation*

As mentioned in Section VI, in this work, we performed two mesoscale simulations. One of these simulations, called WRF-SN, included four-dimensional data assimilation (FDDA) of public-domain observations; the other simulation, called WRF-NN, did not include such data assimilation. In Fig. 21, we present results from the WRF-NN simulation. We also show differences between this simulation and the WRF-SN simulation. It is evident that differences between the WRF-SN and WRF-NN simulations are quite small. This result was not unexpected; since the NARR dataset included a significant amount of observational data, the impact of additional observational data was not significant. However, if one uses other data products with less assimilated data (e.g., NCEP/NCAR Global Reanalysis Project, 2.5-degree resolution, every 6 hours; NCEP GDAS FNL Analysis, 1-degree resolution, every 6 hours; NCEP Eta/NAM AWIP, 40-km resolution, every 6 hours), we would anticipate significantly greater (positive) impact of the use of FDDA in the WRF simulations.

### *2. Grid Resolution*

In addition to the LES runs with 80×80×80 grid points, we performed two additional runs: L64-A-RR-SN, with 64×64×64 grid points and L40-A-R-SN, with 40×40×40 grid points. Results from these runs are presented in Figs. 22–24. The following inferences can be made based on these figures and Figs. 11, 15, and 17:

- The simulated mean profiles are almost insensitive to grid resolution. This is a strength of the LASDD-SGS model and has been reported earlier by Basu and Porté-Agel[8] and Basu et al.[12].

- The strength of the LLJ peak slightly intensifies with increased resolution.

- The resolved variances increase with increasing resolution, as would be expected. However, the difference in the resolved variance is small between the L64-A-R-SN and L80-A-R-SN runs (comparing Fig. 23, right panels, with Fig. 15, right panels). This suggests that the resolved variance may not increase significantly with further enhancement of the grid resolution.

- Inside the boundary layer, the total momentum flux (represented by the local friction velocity) is almost insensitive to grid resolution. However, above the boundary layer, increased resolution helps in dissipating the pile-up of spurious turbulent fluxes.

- The total sensible heat flux values are somewhat sensitive to grid resolution. The surface sensible heat flux values during 7–8 UTC change by ≈25% by changing the resolution from 20 m to 10 m (comparing Fig. 24, bottom left panel, with Fig. 17, bottom right panel). Similar levels of sensitivity of surface sensible heat fluxes to grid resolution were reported by Richardson et al.[73] in the context of idealized SBL simulations.

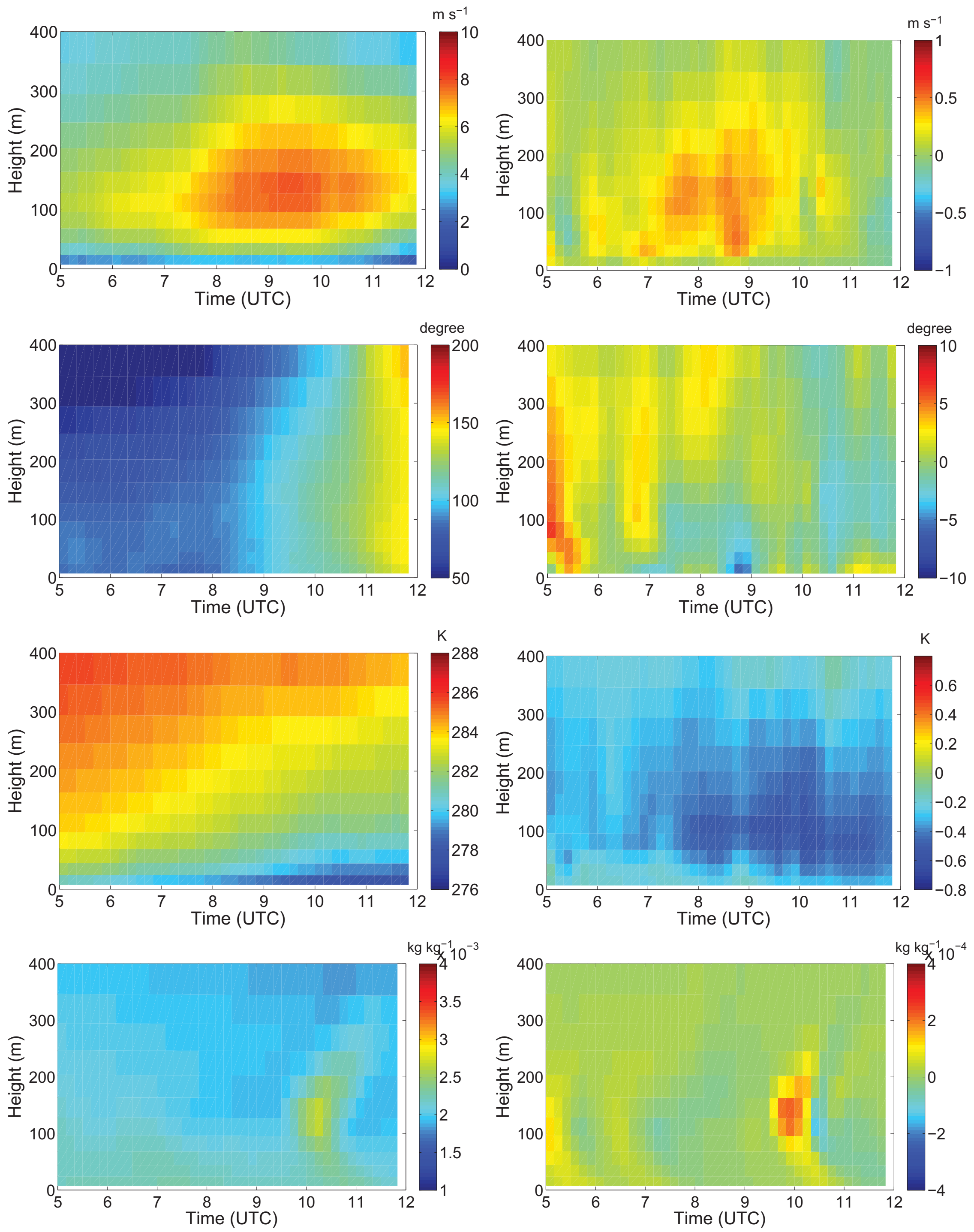

FIG. 21: In the left panel, time-height plots of the WRF model-simulated (domain 4) wind speed (top panel), wind direction (second panel), potential temperature (third panel), and specific humidity (bottom panel) are shown. During this simulation (referred to as the NN simulation), no observational data are assimilated. The right panels show differences between the NN simulation and the control WRF simulation with data assimilation (i.e., NN minus SN).

- As expected, the contribution of the resolved fluxes to the total fluxes increases with increasing resolution. In the L40-A-R-SN run (Fig. 24, bottom right), the resolved sensible heat flux is almost negligible. However, the simulation does not laminarize; the dynamic SGS model essentially acts as a RANS closure in this scenario.

### *3. Longwave Radiation*

In this sub-section, we document the effects of longwave radiation on the LES-generated data. In Fig. 25, results from an LES run with radiation scheme turned off (called L80-A-SN) are presented. In this figure, we also show differences between this simulation and the control simulation (i.e., L80-A-SN minus L80-A-R-SN). The presence of longwave radiational cooling results in a marginally cooler boundary layer and free atmosphere. Near the surface, due to lack of adequate vertical resolution, a small amount of erroneous warming is noticeable. Similar results were reported by Ha and Mahrt[30].

The overall impact of longwave radiational cooling was insignificant for the other variables. Since the geostrophic wind was moderate (see Fig. 8) during the night of October 23/24, radiative flux divergence played a minor role in comparison to turbulent flux divergence.

### *4. Mesoscale Advection*

The influence of mesoscale advection forcing is summarized in Fig. 26. In the left panel of this figure, results from the L80-SN run (with no meoscale advection) are presented. The right panels show differences between this simulation and the control simulation (i.e., L80-SN minus L80-A-R-SN). Since the effects of longwave radiational cooling were found to be marginal, the differences reported in this figure are largely due to the mesoscale advection forcing. The most noteworthy difference is in the height of the stable boundary layer. Due to the lack of thermal advection (see bottom left panel of Fig. 9), the L80-SN run creates a shallower, colder, and more stratified boundary layer in contrast to the L80-A-R-SN run. Due to this stronger stratification, the decoupling of upper air from the surface was stronger than in the control case. This decoupling led to a stronger LLJ, which persisted for much longer than in the control run (Fig. 26).

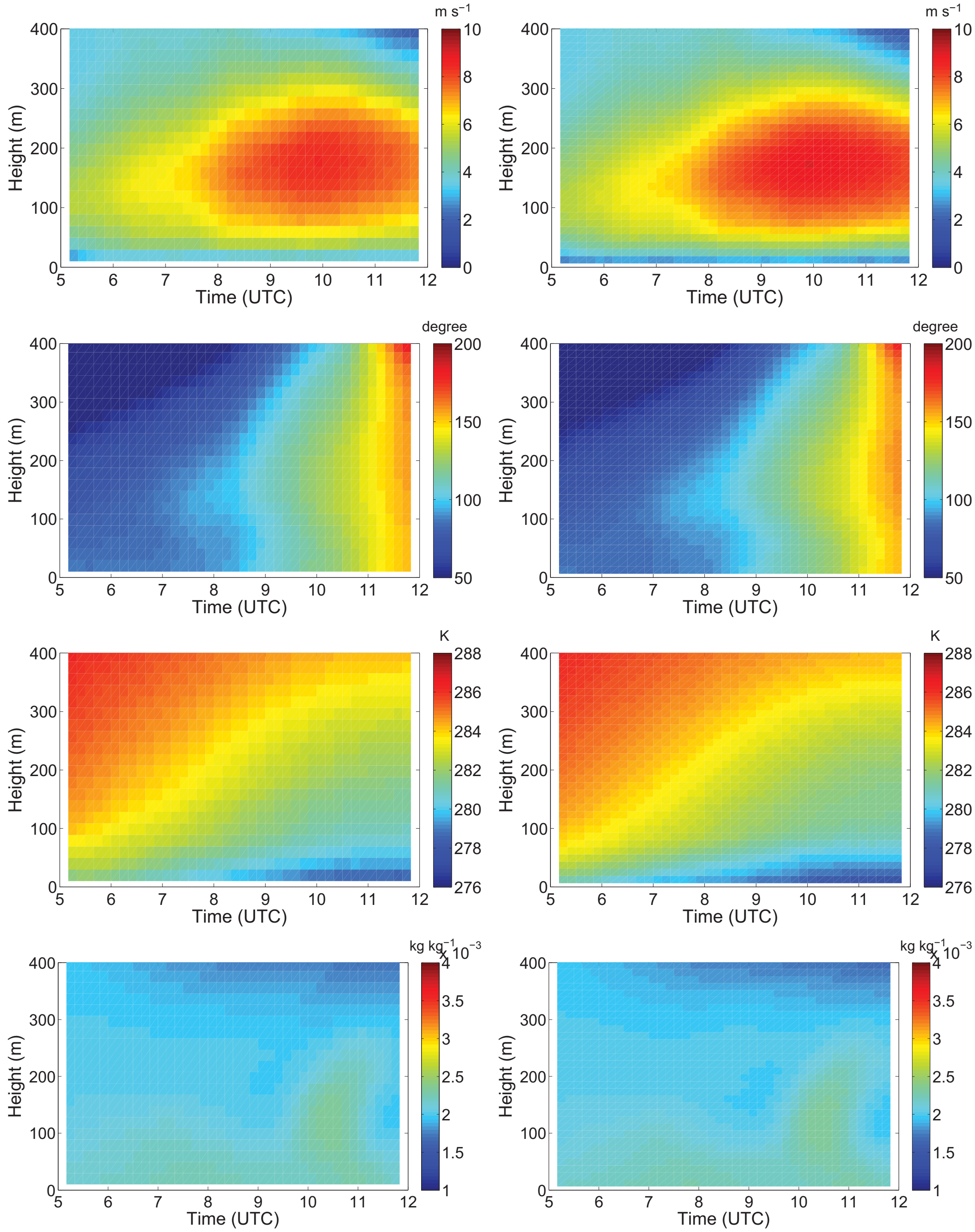

FIG. 22: Time-height plots of simulated wind speed (top panels), wind direction (second panels), potential temperature (third panels), and specific humidity (bottom panels). The left and right panels summarize results from the MATLES model using $40 \times 40 \times 40$ and $64 \times 64 \times 64$ grid points, respectively.

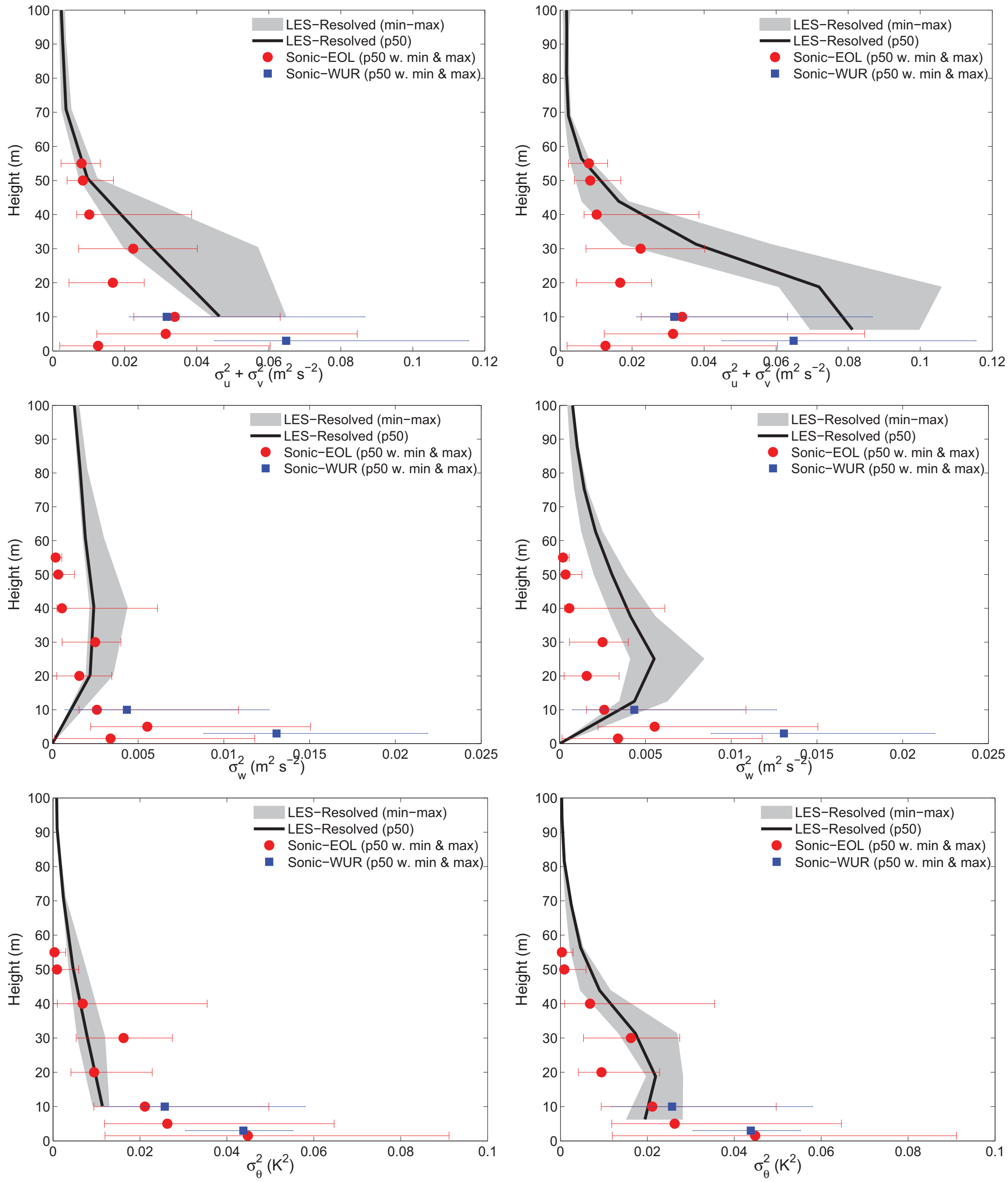

FIG. 23: Plots of variance profiles from the L40-A-R-SN (left panels) and L64-A-R-SN (right panels) runs for 7–8 UTC. The top, middle, and bottom panels represent horizontal velocity variance ($\sigma_u^2 + \sigma_v^2$), vertical velocity variance ($\sigma_w^2$), and potential temperature variance ($\sigma_\theta^2$), respectively. The red dots with whiskers represent median and minimum-to-maximum values of the observations from the 60 m tall meteorological tower. The solid black lines and the light grey areas correspond to the medians and the minimum-to-maximum ranges, respectively, of the LES-generated output data. Note that the simulated results represent resolved variances, whereas the observed data correspond to total variances.

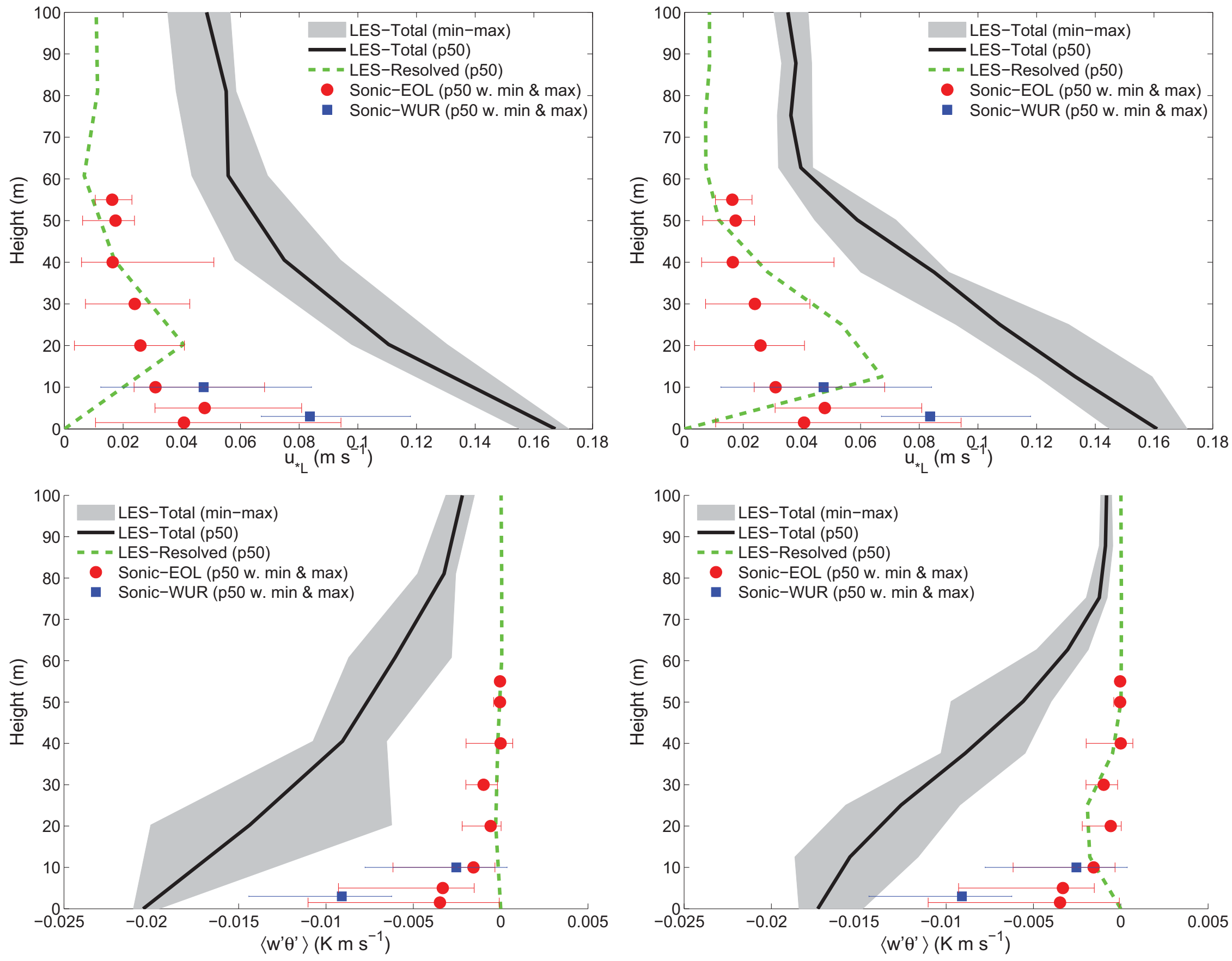

FIG. 24: Plots of local friction velocity (top panels) and sensible heat flux (bottom panels) from the L40-A-R-SN (left panels) and L64-A-R-SN (right panels) runs for 7–8 UTC. The red dots with whiskers represent median and minimum-to-maximum values of the observations from the 60 m tall meteorological tower. The solid black lines and the light grey areas correspond to the medians and minimum-to-maximum ranges of the LES-generated output data, respectively.

## IX. CONCLUDING REMARKS

In this study, we proposed a new coupled mesoscale-large-eddy modeling framework. We demonstrated that if accurate boundary conditions and forcing terms (extracted from a mesoscale simulation) are used, it is possible for a large-eddy simulation to 'mimic' mesoscale model-generated first-order statistics (e.g., hodographs). In this manner, we circumvented any ad-hoc nudging of the LES-generated flow fields.

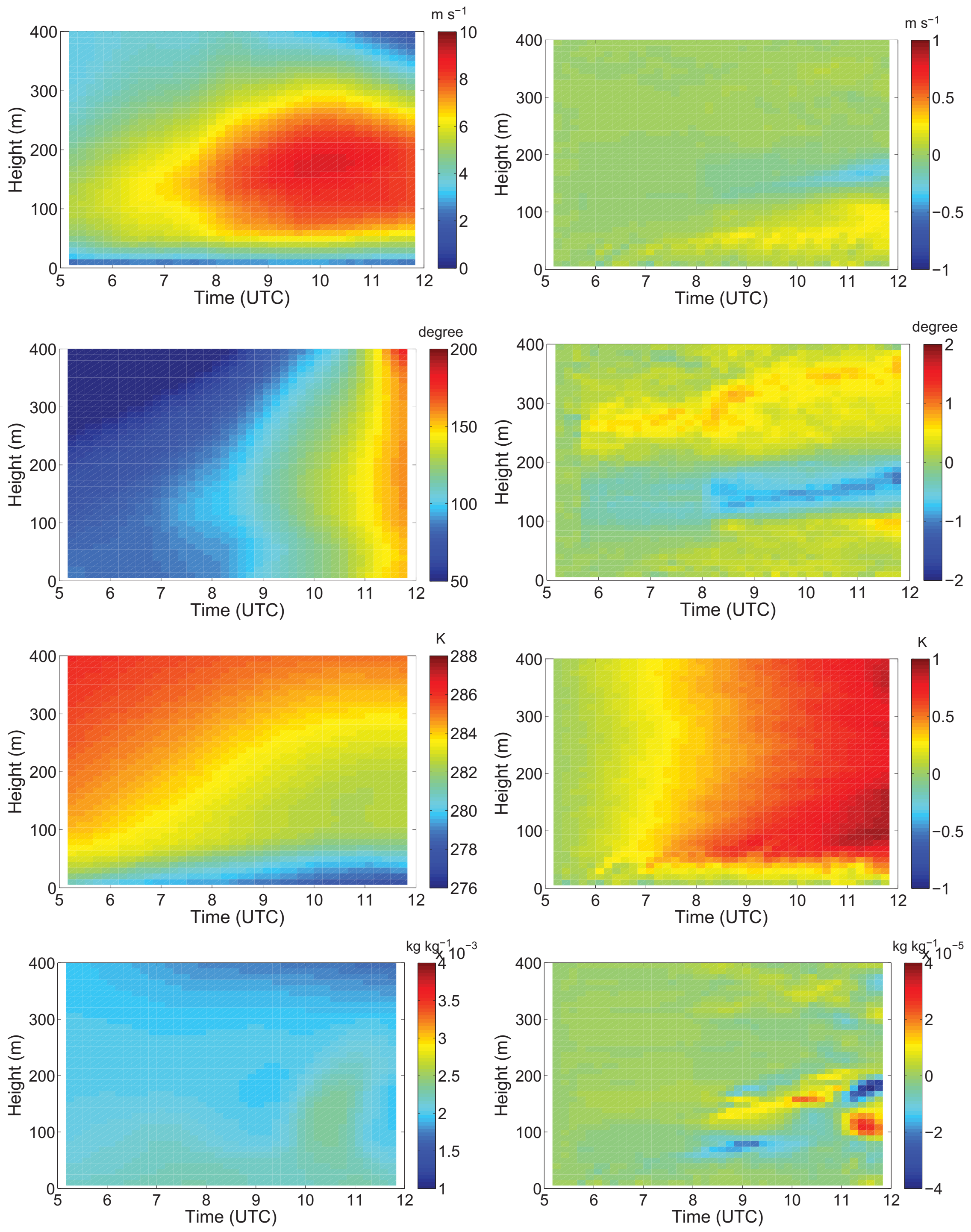

FIG. 25: In the left panels, time-height plots of the MATLES model-simulated wind speed (top panel), wind direction (second panel), potential temperature (third panel), and specific humidity (bottom panel) are shown. The radiation scheme is switched off during this simulation (referred to as L80-A-SN). The right panels summarize differences between this simulation and the control simulation (i.e., L80-A-SN minus L80-A-R-SN).

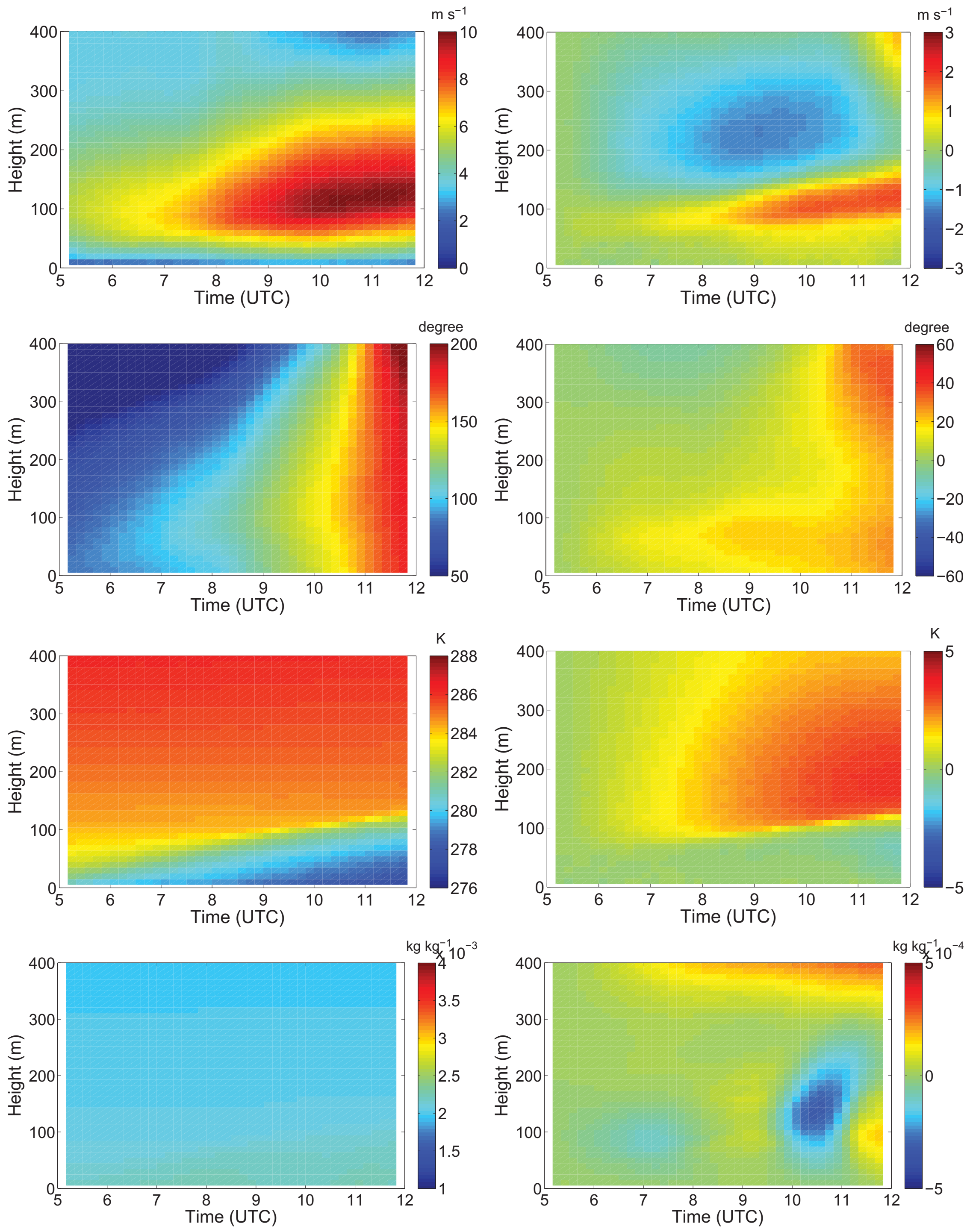

FIG. 26: In the left panels, time-height plots of the MATLES model-simulated wind speed (top panel), wind direction (second panel), potential temperature (third panel), and specific humidity (bottom panel) are shown. During this simulation, the radiation scheme is switched off; also, no mesoscale advection terms are invoked during this simulation (referred to as L80-SN). The right panels summarize difference between this simulation and the control simulation (i.e., L80-SN minus L80-A-R-SN).

The coupled models generated a site-specific realistic stable boundary layer and associated turbulence fields. They reproduced some of the characteristics of an observed low-level jet; the strength of the LLJ, however, was better captured by the LES model than its mesoscale counterpart. More importantly, the LES model, driven by the mesoscale model-generated data, captured different scaling regimes of the energy spectra including the so-called spectral gap. This was one of the key achievements of this study. On the other hand, a major shortcoming of this study is that the coupled models were unable to capture the intermittent nature of the observed surface fluxes.

We found the model validation exercise to be quite challenging. This is due to: (i) unavoidable amplitude and displacement (spatial and temporal) errors associated with mesoscale simulations; (ii) tremendous spatio-temporal variabilities of observed and modeled SBL flow fields; and (iii) significant disagreement among the different observational platforms. This last point was unexpected and needs further emphasis—in this work, we found and reported on disagreements between: lidar and sounding (wind speed); scintillometer and sonic anemometer (surface fluxes); lidar and sonic anemometer (variances). Even two sonic anemometers located a few meters apart provided different data. In closing, we note as Banta[7] also did in highlighting distinctions distinctions between modelers and their models/simulations on the one hand versus observations/data on the other: “everyone believes a measurement except the person who took it,” but “no one believes a model result, except the person who made it.” While such an innocuous statement is generally viewed with some levity, in the context of SBL modeling, we advise modelers not to ‘believe’ based on only one or two sets of observations, but rather validate their simulated results against a diverse suite of observed datasets.

## ACKNOWLEDGMENTS

We are grateful to all those researchers who painstakingly collected data during the CASES-99 field campaign. The authors would like to thank Anantha Aiyyer, Robert Banta, Bert Holtslag, Gary Lackmann, Larry Mahrt, Yelena Pichugina, and Jielun Sun for providing various types of observational data and/or for useful discussions. Computational resources were generously provided by the Renaissance Computing Institute of Chapel Hill, North Carolina. The authors acknowledge the financial support received from the National Science Foundation (AGS-1122315, CBET-0967816, CBET-1050806) and the Air Force Office of Scientific Research (FA9550-12-1-0449). Any opinions, findings and conclusions or recommendations expressed in this material are those of the authors and do not necessarily reflect the views of the National Science Foundation or the Air Force Office of Scientific Research.